\documentclass[aps,prd,showpacs,
twocolumn
,superscriptaddress]{revtex4}
\usepackage{graphicx}% Include figure files
\usepackage{dcolumn}% Align table columns on decimal point
\usepackage{bm}% bold math
\usepackage{color}
\usepackage[normalem]{ulem} % \sout{old text} for strikeout
\usepackage[dvipsnames]{xcolor} % For blue in-text comments
\usepackage{hyperref}
\hypersetup{
  colorlinks=true,        % false: boxed links; true: colored links
  linkcolor=blue,         % color of internal links
  citecolor=red,         % color of links to bibliography
}
\usepackage{mathrsfs}
\usepackage[utf8]{inputenc}
\usepackage{mathtools}
\usepackage{doi}
\usepackage{amsmath}
\usepackage{amssymb}

\begin{document}

\title{Dynamics of Magnetized Particles Around Einstein-\AE ther Black Hole with Uniform Magnetic Field}

\author{Javlon Rayimbaev}
\email{javlon@astrin.uz}
\affiliation{Ulugh Beg Astronomical Institute, Astronomicheskaya 33, Tashkent 100052, Uzbekistan}
\affiliation{Akfa University, Kichik Halqa Yuli Street 17,  Tashkent 100095, Uzbekistan}
\affiliation{National University of Uzbekistan, Tashkent 100174, Uzbekistan}
\affiliation{Institute of Nuclear Physics, Ulugbek 1, Tashkent 100214, Uzbekistan}

\author{Ahmadjon~Abdujabbarov}
\email{ahmadjon@astrin.uz}

\affiliation{Ulugh Beg Astronomical Institute, Astronomicheskaya 33, Tashkent 100052, Uzbekistan}
\affiliation{National University of Uzbekistan, Tashkent 100174, Uzbekistan}
\affiliation{Institute of Nuclear Physics, Ulugbek 1, Tashkent 100214, Uzbekistan}
\affiliation{Tashkent Institute of Irrigation and Agricultural Mechanization Engineers, Kori Niyoziy, 39, Tashkent 100000, Uzbekistan}
\affiliation{Shanghai Astronomical Observatory, 80 Nandan Road, Shanghai 200030, China}

\author{Mubasher Jamil}
\email{mjamil@zjut.edu.cn (corresponding author)}
\affiliation{Institute for Theoretical Physics and Cosmology, Zhejiang University of Technology, Hangzhou 310023, China}
\affiliation{School of Natural Sciences, National University of Sciences and Technology, Islamabad 44000, Pakistan}
\affiliation{Canadian Quantum Research Center 204-3002 32 Ave Vernon, BC V1T 2L7 Canada}

\author{Wen-Biao Han}
\email{wbhan@shao.ac.cn}
\affiliation{Shanghai Astronomical Observatory, 80 Nandan Road, Shanghai 200030, China}

\date{\today}

\begin{abstract}

This work is devoted to study the effects of Einstein-\AE ther gravity on the dynamics of magnetized particles orbiting a static, spherically symmetric and uncharged black hole immersed in an external asymptotically uniform magnetic field in both comoving and proper observers frames. The analysis is carried out by varying the free parameters $c_{13}$ and $c_{14}$ of the Einstein-\AE ther theory and noticing their impacts on the particle trajectories, radii of the innermost stable circular orbits (ISCOs), and the amount of center-of-mass energy produced as a result of the collision. The strength of the magnetic field and the location of circular orbits is significantly affected by varying the above free parameters. We have also made detailed comparisons between the effects of parameters of Einstein-\AE ther and spin of rotating Kerr black holes on ISCO followed by magnetized particles and noticed that both black holes depict similar behaviour for suitable values of $c_{13}$, $c_{14}$, spin and the magnetic coupling parameters which provide exactly the same values for the ISCO. Finally, we have analysed the cases when a static \AE ther black hole can be described as Schwarzschild black hole in modified gravity (MOG) with the corresponding values of the parameters of the black holes.

\end{abstract}

\maketitle

\section{Introduction}

Lorentz invariance is a fundamental consequence and principle of Einstein's special relativity and hence belong to nature itself, which was later brought to a further generalization as diffeomorphism invariance in the general relativity (GR). Due to fundamental limitations of GR to describe physics at the Planck scale and the least understood aspects of quantization of gravity, the assumptions of GR have to be relaxed in order to explore physics near the Planck scale. It is now well-understood that the structure of space near the Planck regime is discrete, obeys non-commutative rules of geometry, violates the Lorentz symmetry, and obeys some form of the generalized uncertainty principle. Among few candidates of Lorentz symmetry violating theories is the Einstein-{\AE}ther theory which is a generally-covariant theory of gravity. In order to violate the Lorentz symmetry, an {\AE}ther field (represented by a timelike vector field) is introduced which is defined by a preferred timelike direction at every point of space \cite{Jacobson01,Eling04}.  In literature, numerous aspects of this theory have been already explored such as cosmological perturbations \cite{Li08,Battye17}, the effects on the generation and propagation of gravitational waves \cite{Han09,Zhang20c}, and the shadow of black holes \cite{Zhu19}, etc. The theory involves several coupling parameters such as $c_{13}$ and $c_{14}$ have been constrained via astrophysical data of the gravitational wave events GW170817 and GRB 170817A \cite{Oost18}. The theory predicts new gravitational-wave polarizations, faster than light propagation speed of gravitational waves without violating causality in some novel ways. Furthermore, by coupling the \AE ther field with the electromagnetic field, two static, electrically charged, and spherically symmetric black hole solutions have been found in the Einstein-{\AE}ther theory \cite{Ding15,Zhang20d}. An n-dimensional extension of charged, static and spherically symmetric black holes is also proposed in this theory \cite{Lin19}. Instead of the Killing horizon, these black holes admit universal horizons. The laws of black hole thermodynamics and the analysis of cosmic censorship conjecture have been studied in \cite{Meiers16}. Recently, one of us explored the phenomenology of the two charged black holes in Einstein-{\AE}ther theory. By investigating the dynamics of a test particle around the black hole in near-circular motion, the properties of quasiperiodic oscillations, epicyclic frequencies, gravitational lensing, periodic orbits, marginally bound orbits and innermost stable circular orbits (ISCO) were recently studied \cite{Azreg20}.  

The Einstein-\AE ther gravity has been used to explore the mechanism to
generate the seed magnetic fields~\cite{Saga13}. These authors have investigated a mechanism of magnetogenesis in the primordial plasma using cosmological
perturbations in the Einstein-\AE ther gravity model where \AE ther field may act as a new source of vector metric perturbations. After the discovery the gravitational waves, the polarization contents of Einstein-\AE ther theory were investigated ~\cite{Gong18}. They have found that there are five polarizations in Einstein-æther theory. These works favor the modification of GR due to the presence of \AE ther may affect different aspects of astrophysical processes \cite{Jacobson08,Jacobson01,Foster06,Garfinkle07}. Here we are interested to study the effects of \AE ther gravity on magnetized particle motion about a static and spherically symmetric black hole characterized by its mass and the parameters of the Einstein-\AE ther theory. 

In this paper, we study the motion of test charged particles around an exact black hole solution in the Einstein-{\AE}ther theory which is surrounded by a test uniform magnetic field. From the astrophysical perspective, the nearby environment of black holes is filled with high energy particles. The evidence to support this claim comes from the electromagnetic spectrum of astrophysical black holes which mainly results from the radiation emission by the particles in the accretion disk and outward collimated jets \cite{Bambi17e}. It is the spacetime geometry of a black hole along side the magnetic field that determines the motion of these particles and the propagation of radiation. The effects of Doppler and gravitational red-shift can be deduced from the electromagnetic spectrum. In this scenario, the study of circular orbits in general and the ISCO, in particular, gives the maximum information about the nearby activity of the black hole. In addition to the gravitational field, the dynamics of particles are moderately affected by the magnetic field in the black hole-accretion disk environment. The origin of the magnetic field around the black hole can be primordial i.e. a relic of the early universe or the gravitational collapse of the dying star carrying the magnetic field \cite{Subramanian16}. In literature, the motion of test charged particles around various kinds of black holes with the uniform magnetic field has been studied extensively \cite{Jawad16,Hussain15,Jamil15,Hussain17,Babar16}. Another important aspect of the black hole accretion disk environment is the particle collisions near the black hole. It was earlier shown that the collision of particles near the black hole horizon could lead to the production of an arbitrarily high center of mass-energy, commonly known as the BSW mechanism \cite{Banados09}. Since then this aspect has received considerable interest from researchers and several aspects of the BSW mechanism have been explored, see \cite{Majeed17,Zakria15}, and references therein. In order to see the effects of the \AE theory parameters more clearly and taking into account the electrical neutrality of most astrophysical back holes, we shall ignore the electric charge parameter in the black hole spacetime.

The black hole cannot have its own magnetic field, however, one may consider the external magnetic field near the black hole. The solution of electromagnetic field equation around the Kerr black hole immersed in an external asymptotically uniform magnetic field was obtained first in Ref.\cite{Wald74}.
In later papers, various properties of electromagnetic field around black hole and neutron stars in external asymptotically uniform dipolar magnetic fields were studied \cite{Aliev86,Aliev89,Aliev02,Frolov11,Frolov12,Benavides-Gallego18,Shaymatov18,Stuchlik14a,Abdujabbarov10,Abdujabbarov11a,Abdujabbarov11,Abdujabbarov08,Karas12a,Stuchlik16,Kovar10,Kovar14,Kolos17,Pulat2020PhRvDMOG,Rayimbaev2019IJMPCS,Rayimbaev2020MPLA,Rayimbaev2019IJMPD}.
In particular, the dynamics of magnetized and charged particle around black holes immersed in external magnetic fields surrounding the magnetically charged black holes have already been studied in diverse gravitational theories \cite{deFelice,defelice2004,Rayimbaev16,Oteev16,Toshmatov15d,Abdujabbarov14,Rahimov11a,Rahimov11,Haydarov20,Haydarov2020EPJC,Abdujabbarov2020PDU,Narzilloev2020PhRvDstringy,Rayimbaev2020PhRvD,TurimovPhysRevD2020,Rayimbaev2020PhysRevDRGI,BokhariPhysRevD2020,Narzilloev2020EPJC1,JuraevaEPJC2021,Abdujabbarov2020Galax,DeLaurentis2018PhRvD,MorozovaV2014PhRvD,Nathanail2017MNRAS,Vrba2020PhRvD,Vrba2019EPJC}. 

This manuscript is organized as follows: In Sec.~\ref{chapter1}, we start with a brief review of Einstein-{\AE}ther black hole immersed in an external magnetic field. Sec.~\ref{chap3} is devoted to studying the magnetized particle motion around Einstein-{\AE}ther black hole in the presence of an external magnetic field. The acceleration process near the Einstein-{\AE}ther black hole is considered in Sec.~\ref{sec4}. We consider the astrophysical applications in Sec.~\ref{appl} and conclude our results in Sec.~\ref{Summary}. 

Throughout this work we use the diagonal metric signature $(-,+,+,+)$ 
for the space-time and geometrized unit system 
$G_N = c = 1$. 
Latin indices run from $1$ to $3$ (or 4 depending on the context)
and Greek ones vary from $0$ to $3$.

\section{Einstein-\AE ther black holes \label{chapter1}}

The action of Einstein-{\AE}ther theory contains the Einstein-Hilbert action with an addition of an action corresponding to a dynamical, unit timelike {\AE}ther field~\cite{Jacobson08,Jacobson01,Foster06,Garfinkle07} which cannot vanish anywhere and breaks local Lorentz symmetry. The complete action has the following form
\begin{eqnarray}\label{actionee}
S_{\ae}=\frac{1}{16\pi G_{\ae}}\int d^4x\sqrt{-g}\left(R+{\cal L}_{\ae}\right),
\end{eqnarray}
where  $g=|g_{\mu \nu}|$ is the determinant of the spacetime metric around a gravitational object in the Einstein-{\AE}ther gravity. The Lagrangian of {\AE}ther field in the action (\ref{actionee}) has the following form:
\begin{eqnarray}
{\cal L}_{\ae}=-M^{\alpha \beta}_{~~~\mu \nu}(D_{\alpha}u^{\mu})(D_{\beta}u^{\nu})+\lambda(g_{\mu \nu}u^{\mu}u^{\nu}+1)\ ,
\end{eqnarray}
here $D_{\alpha}$ is the covariant derivative with respect to $x^\alpha$, $\lambda$ is the Lagrangian multiplier which is responsible for the {\AE}ther four-velocity $u^{\alpha}$ always to be timelike, and $M^{\alpha \beta}_{~~~\mu \nu}$ is defined as
\begin{eqnarray}
M^{\alpha \beta}_{~~~\mu \nu}=c_1g_{\mu \nu}g^{\alpha \beta}+c_2 \delta^{\alpha}_{\mu}\delta^{\beta}_{\nu}+c_3\delta^{\alpha}_{\nu}\delta^{\beta}_{\mu}-c_4 u^{\alpha}u^{\beta} g_{\mu\nu},
\end{eqnarray}
where $c_i$ ($i=1,2,3,4$) are dimensionless coupling constants. Note, that  {\AE}ther gravitational constant has the following form  
\begin{eqnarray}
G_{\ae}=\frac{G_N}{1-\frac{1}{2}c_{14}}\ ,
\end{eqnarray}
where $G_N$ is the Newtonian gravitational constant.

The solution of field equation within the theory~(\ref{actionee}) describing the non-rotating black holes has the following line element in the spherical polar coordinates~\cite{Ding15}:
\begin{eqnarray}\label{metric}
ds^2=-f(r)dt^2+\frac{dr^2}{f(r)}+r^2(d\theta^2+\sin^2\theta d\phi^2)\ ,
\end{eqnarray}
where 
\begin{equation}
f(r)=1-\frac{2 M}{r}\left(1+\frac{2 c_{13}-c_{14}}{4(1-c_{13})}\frac{M}{r}\right) \ ,
\end{equation}
and 
$c_{13}=c_1+c_3$ and $c_{14}=c_1+c_4$, are the new coupling constants of the Einstein-\AE ther theory. 
%%%%%%

Consider the Einstein-\AE ther black hole immersed in an external asymptotically uniform magnetic field. We assume that there exists a magnetic field in the vicinity of black hole which is static, axially symmetric, and homogeneous at the spatial infinity where it has the strength $B_0>0$. The magnetic field is assumed to be weak such that its effect on the spacetime geometry outside the black hole is negligible. In the case when the magnetic field is strong, one needs to modify the spacetime geometry to include the magnetic field. Since the spacetime metric (\ref{metric}) allows timelike and spacelike Killing vectors, one may use Wald's method to find the non-vanishing component of the electromagnetic four potentials \cite{Wald74} 
\begin{eqnarray}\label{Aft}
A_{\phi} & = & \frac{1}{2}B_0r^2\sin\theta.
\end{eqnarray}
The electromagnetic field tensor ($F_{\mu\nu}=A_{\nu,\mu}-A_{\mu,\nu}$) yields the following non-vanishing components
\begin{eqnarray}\label{FFFF}
F_{r \phi}&=&B_0r\sin^2\theta\ , \\
F_{\theta \phi}&=&B_0r^2\sin\theta \cos\theta\ .
\end{eqnarray}
A magnetic field is defined with respect to an observer whose 4-velocity $w_\mu$ defined as follows:  
\begin{eqnarray}
B^{\alpha} =\frac{1}{2} \eta^{\alpha \beta \sigma \mu} F_{\beta \sigma} w_{\mu}\ ,
\end{eqnarray}
where $\eta_{\alpha \beta \sigma \gamma}$ is the pseudo-tensorial form of the Levi-Civita symbol $\epsilon_{\alpha \beta \sigma \gamma}$:
\begin{eqnarray}\label{levi}
\eta_{\alpha \beta \sigma \gamma}=\sqrt{-g}\epsilon_{\alpha \beta \sigma \gamma}\ , \qquad \eta^{\alpha \beta \sigma \gamma}=-\frac{1}{\sqrt{-g}}\epsilon^{\alpha \beta \sigma \gamma}\ ,
\end{eqnarray}
and $g=-r^4\sin^2\theta$. In an orthonormal basis, the magnetic field has the following non-zero components 
\begin{equation}\label{Bzamo}
    B^{\hat{r}}=B_0 \cos\theta, \\ \qquad B^{\hat{\theta}}=\sqrt{f(r)}B_0\sin \theta\ .
\end{equation}

\begin{figure}[h]
    \centering
 \includegraphics[width=0.97\linewidth]{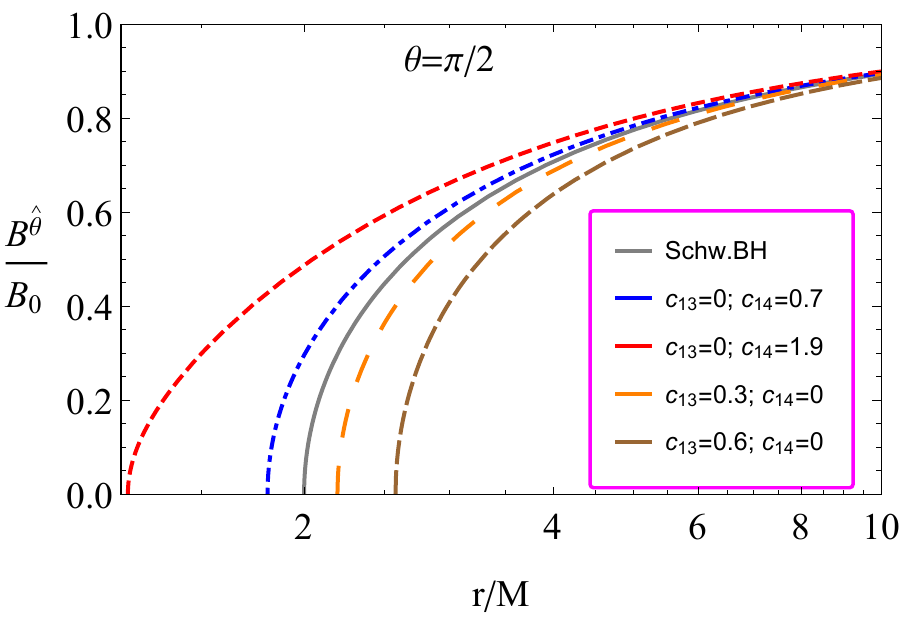}
\caption{Graph showing the radial profile of the normalized angular component of the magnetic field.}
    \label{Bt}
\end{figure}

Figure~\ref{Bt} illustrates the radial dependence of the angular component of magnetic field $B^{\hat{\theta}}$ for different values of parameters $c_{13}$ and $c_{14}$. One can see that the component of the magnetic field varies monotonically with respect to the increase of parameter $c_{14}$ or $c_{13}$.
Moreover, the effects of the \AE ther parameters becomes negligible at far distances and the angular component of external magnetic field takes its asymptotic value. 

\section{Magnetized particle motion in spherically symmetric spacetime\label{chap3}}

 The motion of magnetized particles around any black hole immersed in external magnetic field can be studied using the Hamilton-Jacobi equation 
\begin{eqnarray}\label{HJ}
g^{\mu \nu}\frac{\partial {\cal S}}{\partial x^{\mu}} \frac{\partial {\cal S}}{\partial x^{\nu}}=-\Bigg(m-\frac{1}{2} D^{\mu \nu}F_{\mu \nu}\Bigg)^2\ ,
\end{eqnarray}
where $m$ is the rest mass of the particle and $D^{\mu \nu}F_{\mu \nu}$ represents the interaction between magnetized particle and the external magnetic field.  According to Ref.~\cite{deFelice}, $D^{\alpha \beta}$ can be expressed as 
\begin{eqnarray}\label{DD1}
D^{\mu \nu}=\eta^{\mu \nu \alpha \beta}w_{\alpha }{\cal M}_{\beta} \ , \qquad D^{\alpha \beta }w_{\beta}=0\ ,
\end{eqnarray}
where ${\cal M}^{\alpha}$ is the four-vector of magnetic dipole moment and $w^{\beta}$ is the four-velocity of the particle. The  electromagnetic field tensor $F_{\alpha \beta}$ can be decomposed into electric $E_{\alpha}$ and magnetic $B^{\alpha}$ fields in the following form
\begin{eqnarray}\label{FF1}
F_{\alpha \beta}=w_{\alpha}E_{\beta}-E_{\alpha}w_{\beta}-\frac{1}{2}\eta_{\alpha \beta \sigma \gamma}w^{\sigma}B^{\gamma}\ .
\end{eqnarray}
Now one may easily express the interaction term ${\cal D}^{\mu \nu}F_{\mu \nu}$ in the following form
\begin{eqnarray}\label{DF1}
 D^{\mu \nu}F_{\mu \nu}=2{\cal M} B_0 {\cal L}[\lambda_{\hat{\alpha}}]\ ,
 \end{eqnarray}
where ${\cal M}$ represents the dipolar magnetic moment of the particle and ${\cal L}[\lambda_{\hat{\alpha}}]$ is a function of the space and time coordinates, as well as other parameters defining the tetrad ${\lambda_{\hat{\alpha}}}$ attached to the comoving fiducial observer.

For simplicity here, we consider the orbital motion of magnetized particles around the Einstein-\AE ther black hole in the weak interaction approximation implying $\Big(D^{\mu \nu}F_{\mu \nu}\Big)^2\rightarrow{0}$. We also concentrate our study on the motion of magnetized particles in the equatorial plane, $\theta=\pi/2$ where the angular component of the four-momentum of the particle is zero i.e. $p_{\theta} = 0$. We also consider the magnetic dipole moment of the particle to be perpendicular at the equatorial plane.  The  existence of Killing vectors guarantee the two conserved quantities: $p_{\phi}\equiv L=mu^{\phi}$ and $p_t \equiv -E=mu^{t}$ denoting the total angular momentum and total energy of the particle, respectively. Thus, the Hamilton-Jacobi action for the motion of magnetized particle at the equatorial plane can be splitted as following
\begin{eqnarray}\label{action}
{\cal S}=-E t+L\phi +{\cal S}_r(r)\ .
\end{eqnarray}

One can now easily get the expression for radial motion of the magnetized particle at the equatorial plane by inserting Eq. (\ref{DF1}) and (\ref{action}) in (\ref{HJ}) to get
\begin{eqnarray}\label{rdot}
\dot{r}^2={\cal{E}}^2-V_{\rm eff}(r;c_{13},c_{14},l,{\cal B})\ ,
\end{eqnarray}
where $l=L/(mM)$ and ${\cal E}=E/m$ are specific angular momentum and specific energy of the particle, respectively. The effective potential for radial motion of magnetized particle has the following form 
\begin{eqnarray}\label{effectivepot}
V_{\rm eff}(r;c_{13},c_{14},l,{\cal B})=f(r)\left(1+\frac{l^2}{r^2}-{\cal B} {\cal L}[\lambda_{\hat{\alpha}}]\right)\ ,
\end{eqnarray}
where ${\cal B} = 2 {\cal M} B_0/m$ is the magnetic coupling parameter responsible for interaction between a magnetized particle and the external magnetic field. Here ${\cal B}>0$ (${\cal B}<0$) implies the directions of the external magnetic field and magnetic dipole moment of the particle are the same (opposite), while ${\cal B}=0$ is the case when either there is no external magnetic field or the particle has no magnetic dipole moment. 

In order to estimate the numerical value of $\mathcal B$, let us consider the real astrophysical case of a magnetized neutron star with the typical value of magnetic dipole moment ${\cal M}=(1/2)B_{\rm NS}R_{\rm NS}^3$.
We can estimate the value of the magnetic coupling parameter for the case of the magnetar SGR (PSR) J1745-2900 with the magnetic dipole moment ${\cal M}=1.6 \times 10^{32}\  \rm G \cdot cm^2$ and mass $m\simeq 1.4M$\cite{Mori2013ApJ}, orbiting around Sgr A* as
\begin{equation}
{\cal B}_{\rm PSR J1745-2900}\simeq 0.716 \left( \frac{B_{\rm ext}}{10\ G}\right).
\end{equation}
Now, we shall study the circular orbits using the following standard conditions 
\begin{eqnarray}\label{condition1}
\dot{r}=0 \ , \qquad \frac{\partial V_{\rm eff}}{\partial r}=0\ .
\end{eqnarray}
One may find the magnetic coupling parameter as a function of the radial coordinate, the specific angular momentum and energy, using  expressions (\ref{effectivepot}), (\ref{rdot}) and (\ref{condition1}) in the following form 
\begin{eqnarray}\label{betafunc1}
%&&
{\cal B}(r;l,{\cal E},c_{13},c_{14})=\frac{1}{ {\cal L}[\lambda_{\hat{\alpha}}]}\Bigg(1+\frac{l^2}{r^2}-\frac{{\cal{E}}^2}{f(r)}\Bigg)\ ,
%\\
%&& \frac{\partial V_{\rm eff}}{\partial r}=f(r) {\cal L}[\lambda_{\hat{\alpha}}] \frac{\partial {\cal B}}{\partial r}\ .
  \end{eqnarray} 
The interaction of the magnetized particle and the external magnetic field can be characterised by angular magnetic field and angular dipole moment i.e.
\begin{eqnarray}\label{DF12}
D^{\mu \nu}F_{\mu \nu}=2{\cal M}^{\hat{\theta}}B_{\hat{\theta}}\, .
\end{eqnarray}

The explicit form of ${\cal L}[\lambda_{\hat{\alpha}}]$ can be formulated using the specific tetrad of a fiducial comoving observer. The tetrads for circular motion in the equatorial plane of a Schwarzschild-like black hole reads as \cite{deFelice}
\begin{eqnarray}
\lambda_{\hat{t}}&=&e^{\Psi}\left( \partial_t+\Omega \partial_{\phi} \right)\ , \\\nonumber
\lambda_{\hat{r}}&=&e^{\Psi}\left\{-\frac{\Omega r}{\sqrt{f(r)}}\partial_t-\frac{\sqrt{f(r)}}{r}\partial_{\phi} \right\}\sin(\Omega_{FW} t)\\ &+&\sqrt{f(r)}\cos (\Omega_{FW}t)\partial_r\ , \\
\lambda_{\hat{\theta}}&=&\frac{1}{r}\partial_{\theta}\ , \\\nonumber
\lambda_{\hat{\phi}}&=&e^{\Psi}\left\{\frac{\Omega r}{\sqrt{f(r)}}\partial_t+\frac{\sqrt{f(r)}}{r}\partial_{\phi} \right\}\cos(\Omega_{FW}t)\\
&+&\sqrt{f(r)}\sin(\Omega_{FW}t)\partial_r\ ,
\end{eqnarray} 
where $\Omega_{FW}$ is the Fermi-Walker angular velocity and
\begin{equation}
    \label{epsi}
 e^{\Psi}=\left(f(r)-\Omega^2 r^2\right)^{-\frac{1}{2}}\ ,
\end{equation}
with $\Omega$ denotes the angular velocity of the particles as measured by a distant observer, defined as
\begin{eqnarray}
\Omega=\frac{d\phi}{d t}=\frac{d\phi/d\tau}{d t/d\tau}=\frac{f(r)}{r^2}\frac{l}{{\cal{E}}}\ .
\end{eqnarray}
The components of the magnetic field take the following form
\begin{eqnarray}\label{Bcomp}
&& B_{\hat{r}}=B_{\hat{\phi}}=0\ , \qquad B_{\hat{\theta}}=B_0f(r)\,e^{\Psi}\ ,
\end{eqnarray}
Note that the  orthonormal components of magnetic field in the comoving frame of references in Eq.(\ref{Bcomp}) reduce to the components of the magnetic field as measured by zero-angular-momentum-observer (ZAMO) given in Eq.(\ref{Bzamo}) when $\Omega=0$. 

Now one may calculate the exact form of the interaction part in Hamilton-Jacobi equation by inserting  Eq.~(\ref{Bcomp}) in Eq.~(\ref{DF12}) to get
\begin{eqnarray}\label{DF2}
D^{\mu \nu}F_{\mu \nu}=2{\cal M} B_0f(r)\,e^{\Psi}\ .
\end{eqnarray}
To find the unknown function ${\cal L}[\lambda_{\hat{\alpha}}]$, we  compare the Eqs. (\ref{DF1}) and (\ref{DF2}) which yield
\begin{eqnarray}\label{lambda}
{\cal L}[\lambda_{\hat{\alpha}}]=e^{\Psi}\, f(r)\ .
\end{eqnarray}
Finally, one can calculate the exact form of the magnetic coupling parameter ${\cal B}(r;l,{\cal E},c_{13},c_{14})$  by inserting Eqs. (\ref{lambda}) and (\ref{epsi}) in (\ref{betafunc1}) to obtain
\begin{eqnarray}\label{betafinal}
{\cal B}(r;l,{\cal E},c_{13},c_{14})=\sqrt{\frac{1}{f(r)}-\frac{l^2}{{\cal E}^2 r^2}}\Bigg(1+\frac{l^2}{r^2}-\frac{{\cal E}^2}{f(r)}\Bigg)\ .
\end{eqnarray}

The above equation has the following  physical meaning: a magnetized particle with specific energy ${\cal E}$ and angular momentum $l$ can be found in the circular orbit at a certain distance $r$ from the central object with the corresponding value of the magnetic interaction parameter which can be calculated from Eq.~(\ref{betafinal}).

\begin{figure}
  \centering
   \includegraphics[width=1\linewidth]{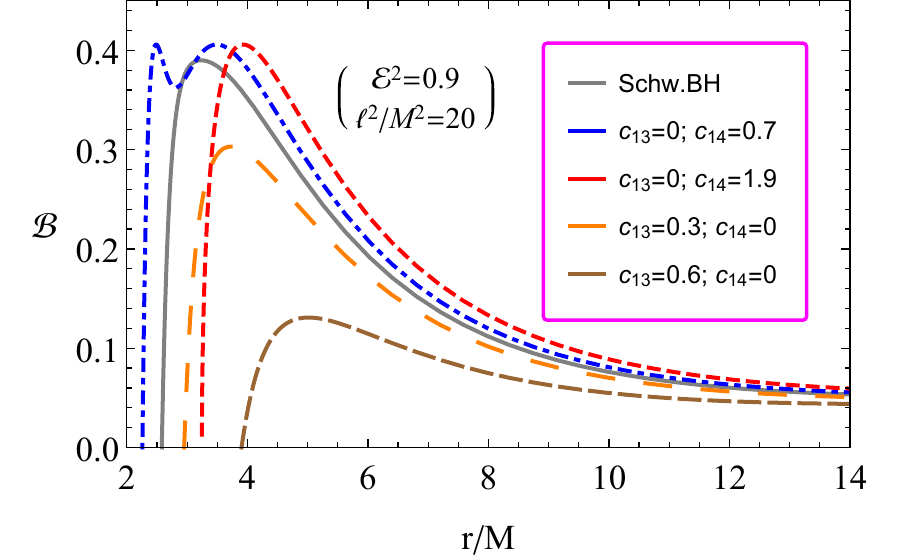}
   \includegraphics[width=1\linewidth]{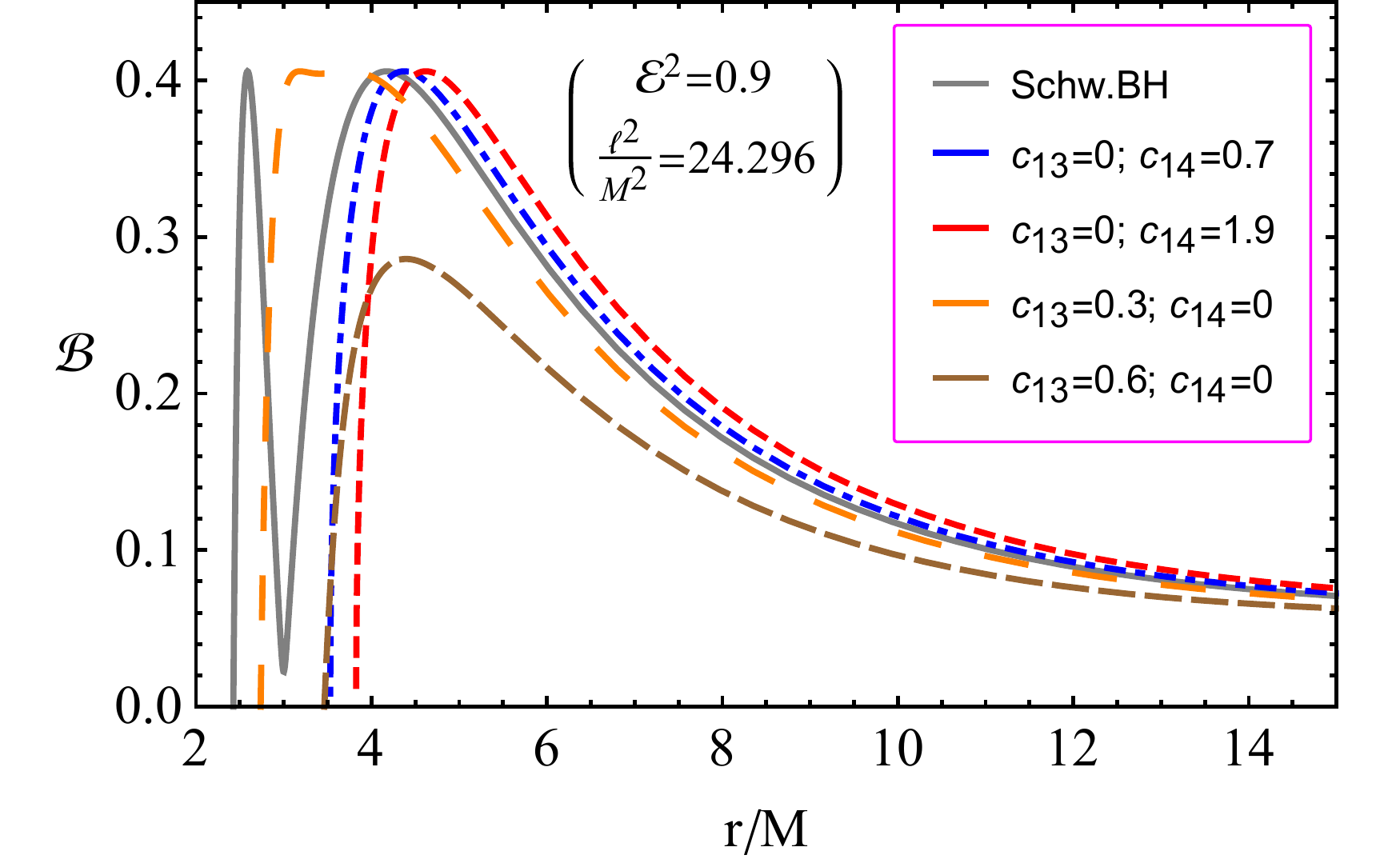}
   \caption{The radial profile of magnetic coupling function for different values of the specific angular momentum: $l^2/M^2=20$ (top panel) and $l^2/M^2=24.296$ (bottom panel) with ${\cal E}^2=0.9$. \label{betafig}}
\end{figure}

The radial profile of magnetic coupling function $\mathcal B$ for different values of $c_{13}$ and $c_{14}$ parameters is depicted in Fig.~\ref{betafig}. One observes that by the increase of the parameter $c_{14}$ ($c_{13}$) leads to an increase (decrease) in the maximum magnetic coupling parameter.

We now start to analyse the behavior of magnetic coupling parameter corresponding to the stable circular orbits. The conditions for the stable circular orbits for magnetized particles have the following form 
\begin{eqnarray}\label{betacondition}
{\cal B}={\cal B}(r;l,{\cal E},c_{13},c_{14}), \quad \frac{\partial {\cal B}(r;l,{\cal E},c_{13},c_{14})}{\partial r}=0\ .
\end{eqnarray}
This is a system of two equations with six unknown quantities
${\cal B},r,l,{\cal E}, c_{13},c_{14}$, hence its solution
can be parameterized in terms of any two among the five independent
variables. Here we use the magnetic coupling ${\cal B}$ and the orbital radius $r$ as free parameters. Our aim is then to find the angular momentum $l$ and the specific energy ${\cal E}$ of
the particle as functions of $r$ and ${\cal B}$. 
First, one can find the ``minimum energy" of the particle which corresponds to the minimum value of magnetic interaction parameter, and is determined using the second condition (\ref{betacondition}) and thereby solving it for the specific energy 
\begin{eqnarray}\label{emin}
&&{\cal E}_{\rm min}(r;l,c_{13},c_{14})=\\\nonumber &&= \frac{l\left[\left(1-\frac{2M}{r}\right)+\frac{c_{14}}{2}\frac{M^2}{r^2}-c_{13} \left(1-\frac{M}{r}\right)^2\right]}{\sqrt{\left(1-c_{13}\right) M \left[r-c_{13} (r-M)-\frac{c_{14}}{2} M\right]}}\ .
\end{eqnarray}

\begin{figure}[h!]
  \centering
   \includegraphics[width=0.95\linewidth]{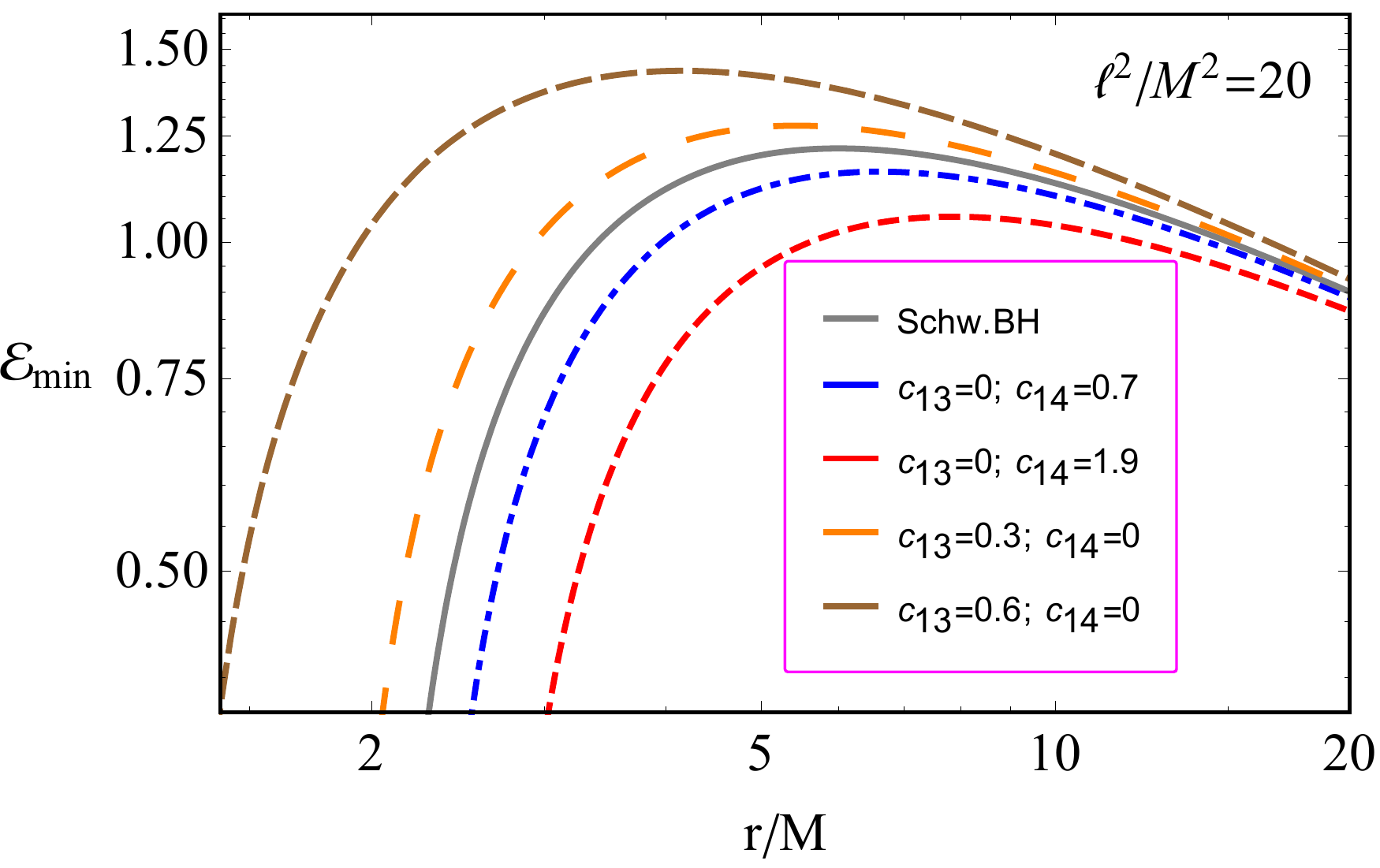}
      \caption{The dependence of the minimum value of specific energy of the magnetized particles  on radial coordinate. \label{eminfig}}
\end{figure}

The radial profile of the minimum value of specific energy of magnetized particle is presented in Fig.~\ref{eminfig} by varying the parameters of Einstein-\AE ther gravity. An increase in the parameter $c_{13}$ ($c_{14}$) causes an increase (decrease) in the maximum value of the specific energy. Moreover, the distances where stable orbits exist and the energy takes its maximum shift to the central object  due to the increase of the parameter  $c_{13}$ ($c_{14}$). It implies that  $c_{13}$ plays a role of a source of an additional gravity, while $c_{14}$ decreases the gravitational potential of the central black hole.

The minimum of the magnetic coupling parameter can be found by substituting Eq.~(\ref{emin}) in  Eq.~(\ref{betafinal})
we have
%\begin{widetext}
\begin{eqnarray}\label{betamineq}\nonumber
&&{\cal B}_{\rm min}(r;l,c_{13},c_{14})=\frac{2 \sqrt{1-c_{13}}}{M r \left[2 c_{13} (r-M)+c_{14} M-2 r\right]}
\\\nonumber
&& \times \frac{\sqrt{c_{14} M^2+r (r-3 M)-c_{13} \left(2 M^2-3 M r+r^2\right)}}{c_{14} M^2-2 c_{13} (r-M)^2+2 r (r-2 M)} \\\nonumber
&& \times \Big\{c_{14} M^2 \left(2 l^2+r^2\right)-2 \left[r-c_{13} (r-M)\right]\\
&&\times  \left[M r^2-l^2 (r-3 M)\right]\Big\}\ .
\end{eqnarray}
%\end{widetext}

\begin{figure}
  \centering
   \includegraphics[width=0.98\linewidth]{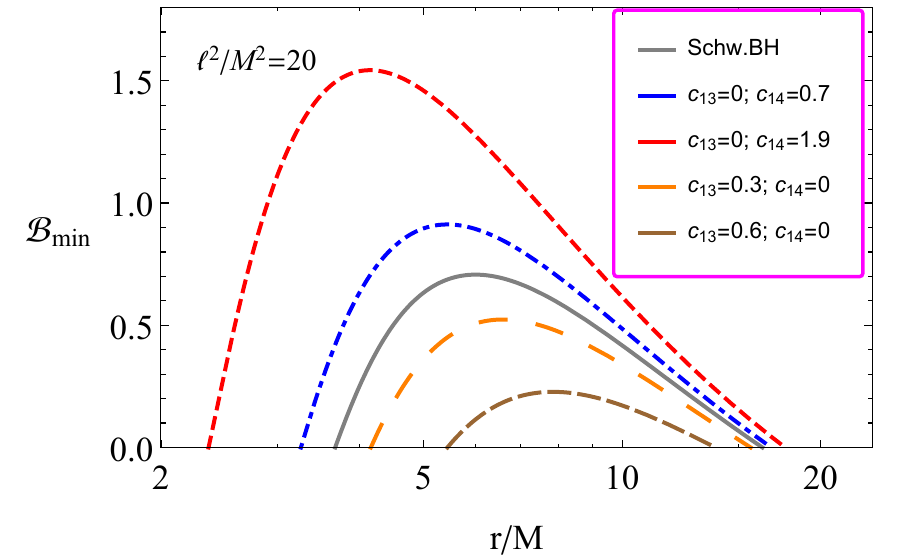}
    \caption{The dependence of the minimum value of the magnetic coupling parameter on the radial coordinate for the fixed value of the specific angular momentum of the particle $l^2/M^2=20$. \label{betaminfig}}
\end{figure}

Figure~\ref{betaminfig} shows the radial dependence of the minimal value of magnetic coupling parameter of the magnetized particle for different values of $c_{13}$ and $c_{14}$ and the fixed value of the specific angular momentum of the magnetized particles.  As we showed in Fig.~\ref{eminfig} the parameter $c_{14}$ decreases the gravity of the central black hole, therefore the effects of magnetic field becomes dominant. Hence the magnetic interaction helps the magnetized particles to stay in circular orbits together with the effect due to gravitational forces.  

The upper limit for the radius of stable circular orbit can be calculated using the extreme value of the magnetic coupling parameter which corresponds to some minimum value of the specific angular momentum. One may obtain the expression for the angular momentum by solving the equation $\partial {\cal B}_{\rm min}/\partial r=0 $ and solving for $l$ gives 
\begin{eqnarray}\label{lmineq}
%\nonumber
&&l_{\rm min}^2(r;c_{13},c_{14})=\frac{M^2 r^2}{2}\\\nonumber && \times \frac{\left[2 c_{13} (r-M)+c_{14} M-2 r\right]^2}{2 c_{13} (r-M) (2 r-M)+6 M r-4 r^2-c_{14} M^2}
\\\nonumber
&& \times \Big[c_{13} \left(2 M^2-3 M r+r^2\right)-c_{14} M^2-r (r-3 M)\Big]^{-1}.
\end{eqnarray}

\begin{figure}[h!]
  \centering
   \includegraphics[width=0.98\linewidth]{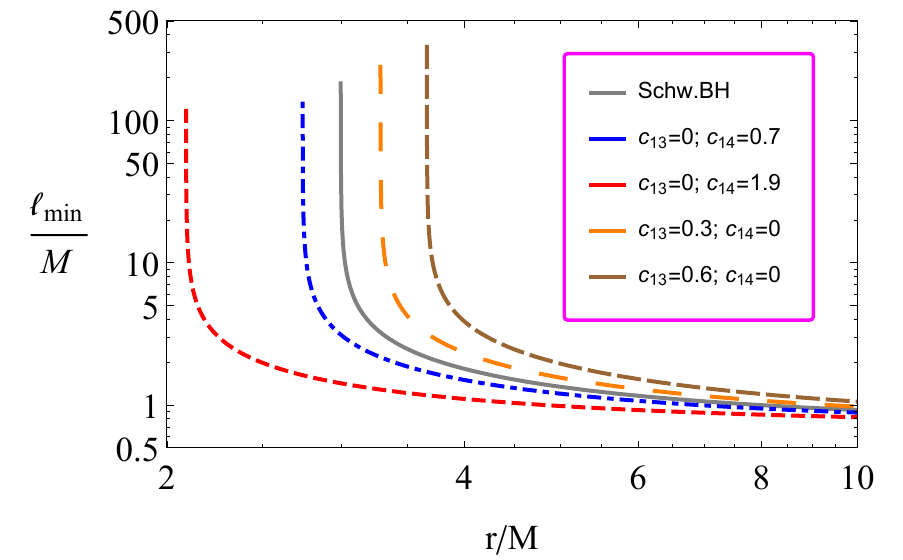}\label{lminfig}
    \caption{The radial profile  of minimal value of the specific angular momentum of the magnetized particles around an \AE ther black hole corresponding to stable circular orbits. }
\end{figure}

Figure \ref{lminfig} shows the radial dependence of the minimum specific angular momentum for different values of  $c_{13}$  and $c_{14}$. One can see from the figure that the maximum value of the minimum angular momentum increase (decrease) with the increase of the parameter $c_{13}$ ($c_{14}$). However, the distance where the specific angular momentum takes the maximum increases (decreases) with increase of the parameter $c_{14}$ ($c_{13}$).

Now it is possible to obtain and analyze the extreme value for the magnetic interaction parameter substituting Eq.~(\ref{lmineq}) into Eq.~(\ref{betamineq}) and we have
\begin{eqnarray}
&&{\cal B}_{\rm extr}(r;c_{13},c_{14})=4r \sqrt{1-c_{13}}
\\\nonumber
&& \times \frac{\sqrt{c_{13} \left(3 M r-2 M^2-r^2\right)+c_{14} M^2+r (r-3 M)}}{2 c_{13} (r-M) (2 r-M)+6 M r-4 r^2-c_{14} M^2}\ .
\end{eqnarray}

\begin{figure}[h!]
  \centering
   \includegraphics[width=0.98\linewidth]{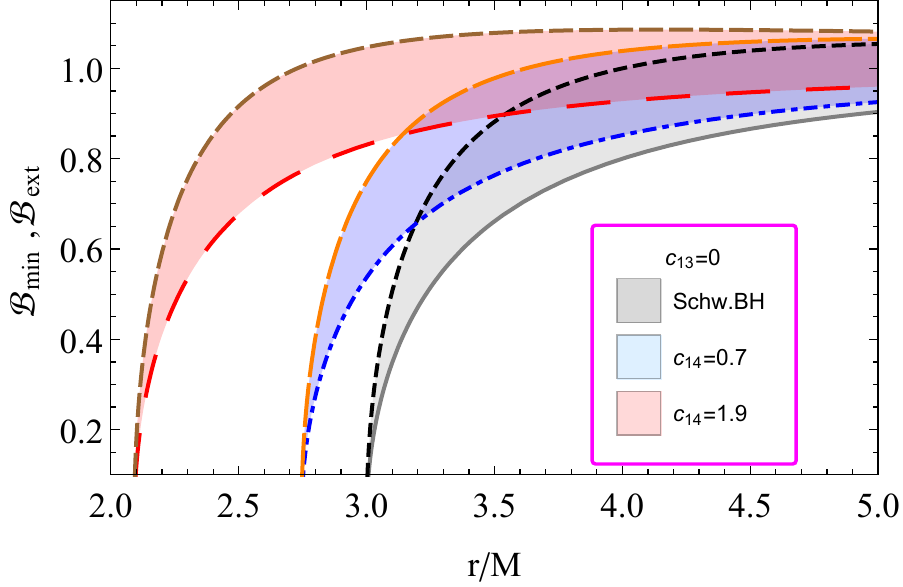}
   \includegraphics[width=0.98\linewidth]{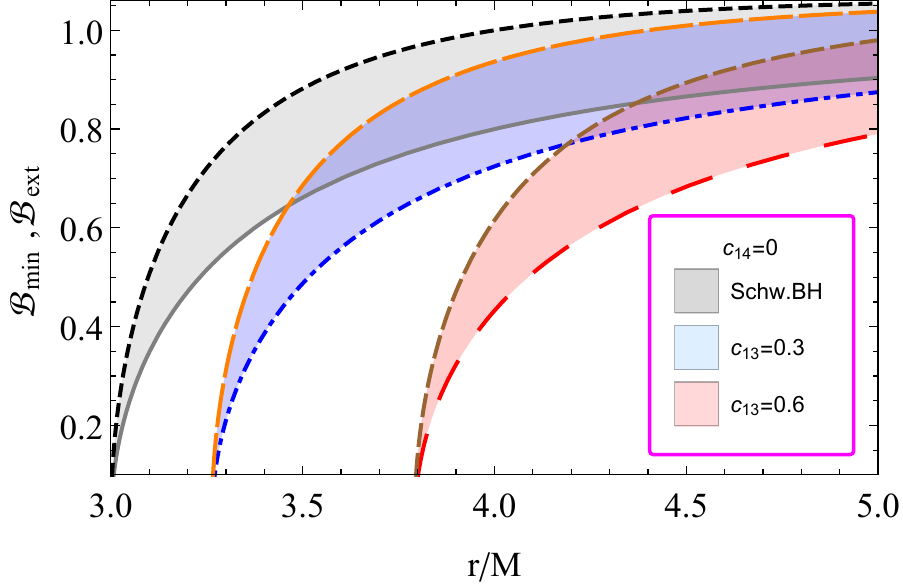}
    \caption{The radial dependence of minimal and extreme values of magnetic coupling parameter. The colored are between the minimal (upper curve lines) and extremal values (lower curve lines) of the coupling parameter implies the range of the values of the magnetic coupling parameter at a given radius the particles orbits to be stable and circular.}     \label{betaextremfig}
\end{figure}

Figure~\ref{betaextremfig} illustrates the radial dependence of the extreme value of the magnetic coupling parameter and the minimum value of the magnetic coupling parameter at $\mathcal B_{\rm min}(l=0)$, for the different values of the parameters $c_{13}$ and $c_{14}$. The range where stable circular orbits allowed for the magnetized particle with the magnetic coupling parameter $\mathcal B_{\rm extr}<\mathcal B<\mathcal B_{\rm min}(l=0)$ is shown with the colored areas. One can see from the Fig.~\ref{betaextremfig} that the minimum distance of the circular orbits increases (decreases) with the increase of the values of the parameter $c_{13}$ ($c_{14}$). The range of stable circular orbits (between $r_{\rm min}$ and $r_{\rm max}$) for the magnetized particles is illustrated by the colored areas in Fig.\ref{betaextremfig} and can be found by solving the equations ${\cal B}_{\rm min}(l=0) = {\cal B} $ and ${\cal B}_{\rm extr} = {\cal B} $, with respect to the radial coordinate, respectively. The extreme value of the magnetic coupling parameter grows by increasing the radial coordinate and it tends to one at when $r \to \infty$. It implies that there are no stable circular orbits for magnetized particles with ${\cal B}>1 \geq {\cal B}_{\rm extr}$ around black holes immersed in an external asymptotically uniform magnetic field. One may conclude that the magnetic coupling parameter has to be $1>{\cal B}>{\cal B}_{\rm extr}$.  In the astrophysical observations of magnetized particles such as recycled radio pulsars and/or magnetars around supermassive black holes at the center of galaxies can be estimated upper limits for the values of the external magnetic field where circular stable orbits of magnetized objects take place by using these assumptions.

 The effects of Einstein-\AE ther gravity parameters over the range where circular orbits are allowed ($\Delta r = r_{\rm max}-r_{\rm min}$) are shown in Table~\ref{tab} for the fixed magnetic coupling ${\cal B}=0.1$ and the different values of the parameter of Einstein-\AE ther gravity. The range of $\Delta r$ is given in the unit of {$1.5(M/M_{\odot})$ meters}. 

\begin{table}[h!] \begin{center}\begin{tabular}{|c| c| c| c| c|}\hline
$ $ &  $c_{14}=0$ & $c_{14}=0.1$ & $c_{14}=0.5$ & $c_{14}=0.8$ \\[1.5ex]\hline \hline
$c_{13}=0$ & $-$ & 1.1942 & 1.8717 & 3.5971 \\[1.5ex] \hline
$c_{13}=0.1 $ & 0.9197 & 1.1787 & 1.7995 & 3.4756 \\[1.5ex]\hline
$c_{13}=0.5$ & 0.7928 & 0.8679 & 1.4951 & 2.9577 \\[1.5ex]\hline
$c_{13}=0.8$ & 0.6417 & 0.7108 & 1.2564& 2.527 \\[1.5ex]\hline
\end{tabular} \end{center}
\caption{\label{tab} Numerical values for $\Delta r=r_{\rm max}-r_{\rm min}$ in the unit of {$1.5(M/M_{\odot})$ meters} with  ${\cal B}=0.1$ for the different values of the parameters of Einstein-\AE ther theory.}
 \end{table}

 One can see from Table.\ref{tab} that the increase of the parameter $c_{13}$ cause increasing of the range $\Delta r$, while with increasing the parameter $c_{14}$ the range narrows. In the case of real astrophysical scenarios of the observations of orbits of hot spots around Sgr A*, assuming the hot spots are magnetized objects, one may estimate the magnetic coupling parameter for the "hot-spot"- objects using the observational data of the measurements of the difference of their orbits.

\section{Acceleration of particles near the Einstein-{\ae}ther black hole \label{sec4}}

In this section, we will study the collision of particles near the black hole in Einstein-\AE ther gravity. Particularly, we consider the center of mass energy of two particles near the black hole immersed in the magnetic field. We consider the effect of black hole parameters and external magnetic field to study the center-mass-energy for the two colliding particles coming from infinity with energies ${\cal E}_1$ and ${\cal E}_2$. The center of mass energy of two particles can be found using the expression ~\cite{Banados09}
\begin{eqnarray}\label{cmenergy}
{\cal E}_{cm}^2=\frac{E_{cm}^2}{2m_0c^2}=1-g_{\alpha \beta}u_1^{\alpha}u_2^{\beta} \ , 
\end{eqnarray}
where $u_1^{\alpha}$ and $u_2^{\beta}$ are four-velocities of the colliding particles. Below we investigate head-on collisions of magnetized particles with magnetized, charged, and neutral particles at the equatorial plane where $\theta=\pi/2$ with the initial energies ${\cal E}_1={\cal E}_2=1$.

\subsection{Two magnetized particles}
First, we consider the case of two magnetized particles collision. The four-velocity of the magnetized particle at the equatorial plane has the following non-zero components:
\begin{eqnarray}\label{4vel1}
\nonumber
\dot{t}&=&\frac{{\cal E}}{f(r)}\ ,
\\
\nonumber
\dot{r}^2&=&{\cal E}^2-f(r)\left[\left(1-{\cal B} \sqrt{f(r)} \right)^2+\frac{l^2}{r^2}\right]\ ,
\\
\dot{\phi}&=&\frac{l}{r^2}\ .
\end{eqnarray}

The expression for center of mass energy for the two magnetized particles can be determined by substituting four-velocities Eqs. (\ref{4vel1}) into (\ref{cmenergy}) in the following form
\begin{eqnarray}\label{cmenergy1}\nonumber 
{\cal E}_{\rm cm}^2&=&1+\frac{{\cal E}_1{\cal E}_2}{f(r)}-\frac{l_1l_2}{r^2}-\\
\nonumber
&-&\frac{1}{f(r)}\sqrt{{\cal E}_1^2-f(r)\left[\left(1-{\cal B}_1 \sqrt{f(r)} \right)^2+\frac{l_1^2}{r^2} \right]}
\\
&\times & \sqrt{{\cal E}_2^2-f(r)\left[\left(1-{\cal B}_2 \sqrt{f(r)} \right)^2+\frac{l_2^2}{r^2}\right]}\ . 
\end{eqnarray}
%%

%%%%%%%%%%%%%%%
\begin{figure}[h!]
    \centering
\includegraphics[width=0.98\linewidth]{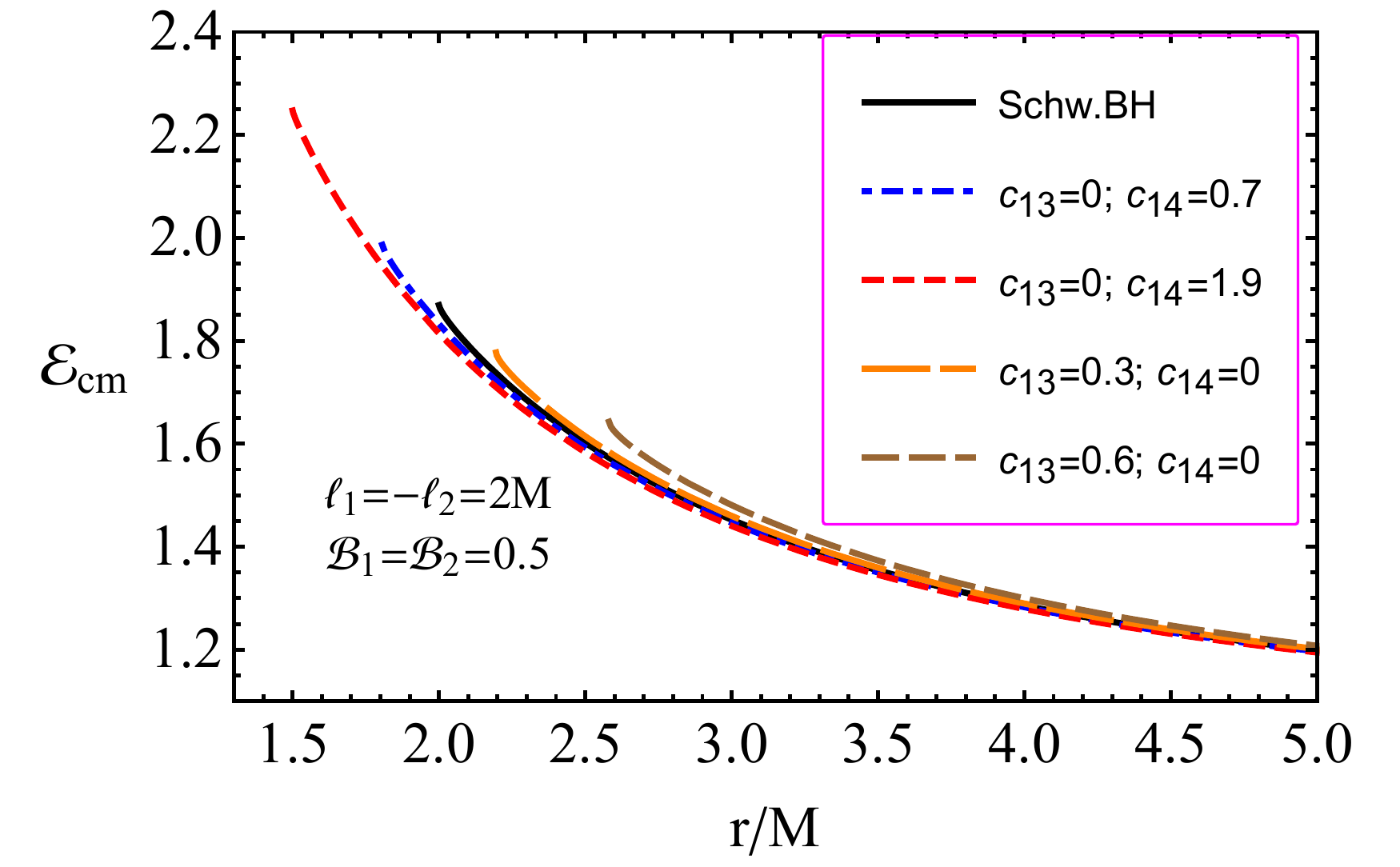}
    \includegraphics[width=0.98\linewidth]{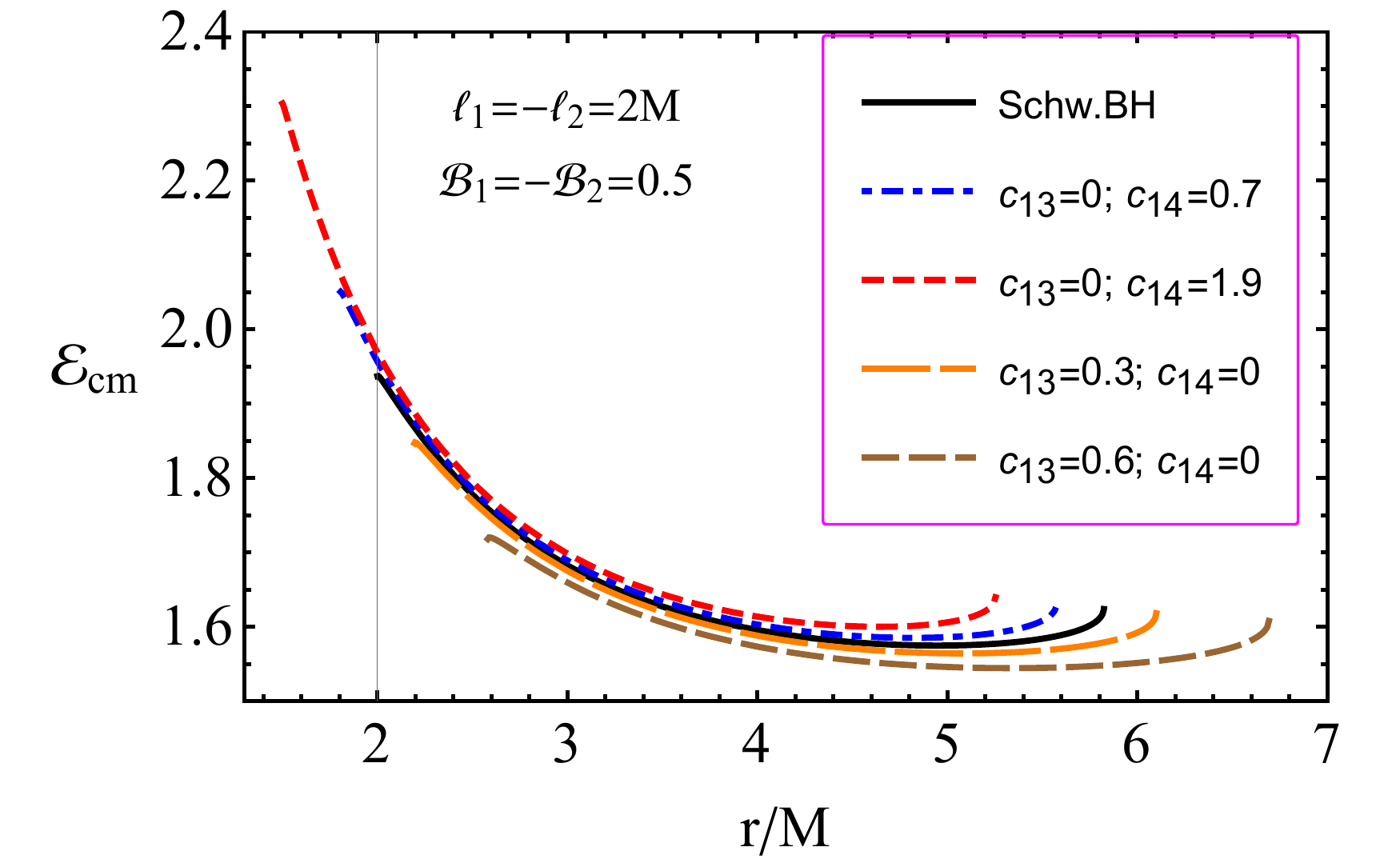}
       \caption{The radial dependence of the center-of-mass energy for the collisions of two magnetized particles with positive-positive (top panel) and positive-negative (bottom panel) values of the magnetic coupling parameter around a static black hole in Einstein-\AE ther gravity along with comparison of the Schwarzschild case. In both cases, particles have same values of the specific angular momentum: $l_1=-l_2=2M$.}
    \label{centermm}
\end{figure}

Figure~\ref{centermm} shows the radial dependence of center-of-mass energy of head-on collision of two magnetized particles with the magnetic coupling parameter ${\cal B}_1={\cal B}_2=0.1$ (on the top panel) and ${\cal B}_1=0.1; {\cal B}_2=-0.1$ around static black holes immersed in the external magnetic field in Einstein - \AE ther gravity. We assume specific angular momentum of the particles with values: $l_1/M=2$ and $l_2/M=-2$. It shows that the collisions of magnetized particles with the same directions as their magnetic dipole moment increases (decreases) the center-of-mass energy by the increase of the parameter $c_{14}$ ($c_{13}$). However, in the case when the directions of the dipoles are opposite, the center of mass energy of the collisions of magnetized particles disappears far from the central object due to different direction of the momentum of colliding magnetized particles with different orientation of proper magnetic dipoles. It implies that where the energy disappeared the collision does not take place due to the dominated repulsive behaviour of the interaction between the magnetized particles. It is shown that the maximal distance where the collision of the magnetized particles takes place increase (decreases)  with the increase of the \AE ther parameter $c_{13}$ ($c_{14}$). Moreover, one can see from the bottom panel that the value of center-of-mass energy of the colliding particles cuts out at a critical distance $r_{cr}$ depending on the value of the \AE ther parameters. It implies that at the distances $r>r_{cr}$ the collisions of the magnetized particle does not occur if their magnetic dipoles are opposite to each other.

\subsection{ Magnetized and charged particles}

In this subsection we consider collision of magnetized and charged particles. Four-velocities for charged particles can be found using the Lagrangian for the charged particles  with the electric charge $e$ and mass $m$ in the presence of electromagnetic field 
\begin{eqnarray}
 \mathscr{L}=\frac{1}{2}mg_{\mu \nu}u^{\mu} u^{\nu}+e u^{\mu}A_{\mu}\ .
\end{eqnarray}
The conserved quantities: the energy and the angular momentum can be found as
\begin{eqnarray}
&&p_t=\frac{\partial \mathscr{L}}{\partial \dot{t}}=mg_{tt}\dot{t}\, ,
\\
&&p_{\phi}=\frac{\partial \mathscr{L}}{\partial \dot{\phi}}=mg_{\phi \phi}\dot{\phi}+eA_{\phi},
\end{eqnarray}
and four-velocity of the charged particle at equatorial plane has the following nonzero components 
\begin{eqnarray}\label{4vel2}
\nonumber
\dot{t}&=&\frac{{\cal E}}{f(r)}\ ,
\\
\nonumber
\dot{r}^2&=&{\cal E}^2-f(r)\Bigg[1+\Bigg(\frac{l}{r}-\omega_{\rm B} r\Bigg)^2\Bigg]\ ,
\\
\dot{\phi}&=&\frac{l}{r^2}-\omega_{\rm B}\ ,
\end{eqnarray}
where $\omega_B=e B/(2mc) $ is the cyclotron frequency which is responsible for the interaction the external magnetic field and charged particle. One may get the expression for the center-of-mass energy of the colliding charged and magnetized particles from substituting four-velocities of the particles Eqs. (\ref{4vel1}) and (\ref{4vel2}) into Eq.(\ref{cmenergy}) and it takes the form:
\begin{eqnarray}\label{cmenergy2}
\nonumber
{\cal E}_{\rm cm}^2&=&1+\frac{{\cal E}_1{\cal E}_2}{f(r)}-\left(\frac{l_1}{r^2}-\omega_{\rm B}\right)l_2\\
\nonumber
&-&\frac{1}{f(r)}\sqrt{{\cal E}_1^2-f(r)\left[1+\Big(\frac{l_1}{r}-\omega_{\rm B} r\Big)^2\right]}
\\
& \times &  \sqrt{{\cal E}_2^2-f(r)\left[\left(1-{\cal B} \sqrt{f(r)} \right)^2+\frac{l_2^2}{r^2}\right]}\ .
\end{eqnarray}

%%%%%%%%%%%%%%%
\begin{figure}
    \centering
  \includegraphics[width=0.98\linewidth]{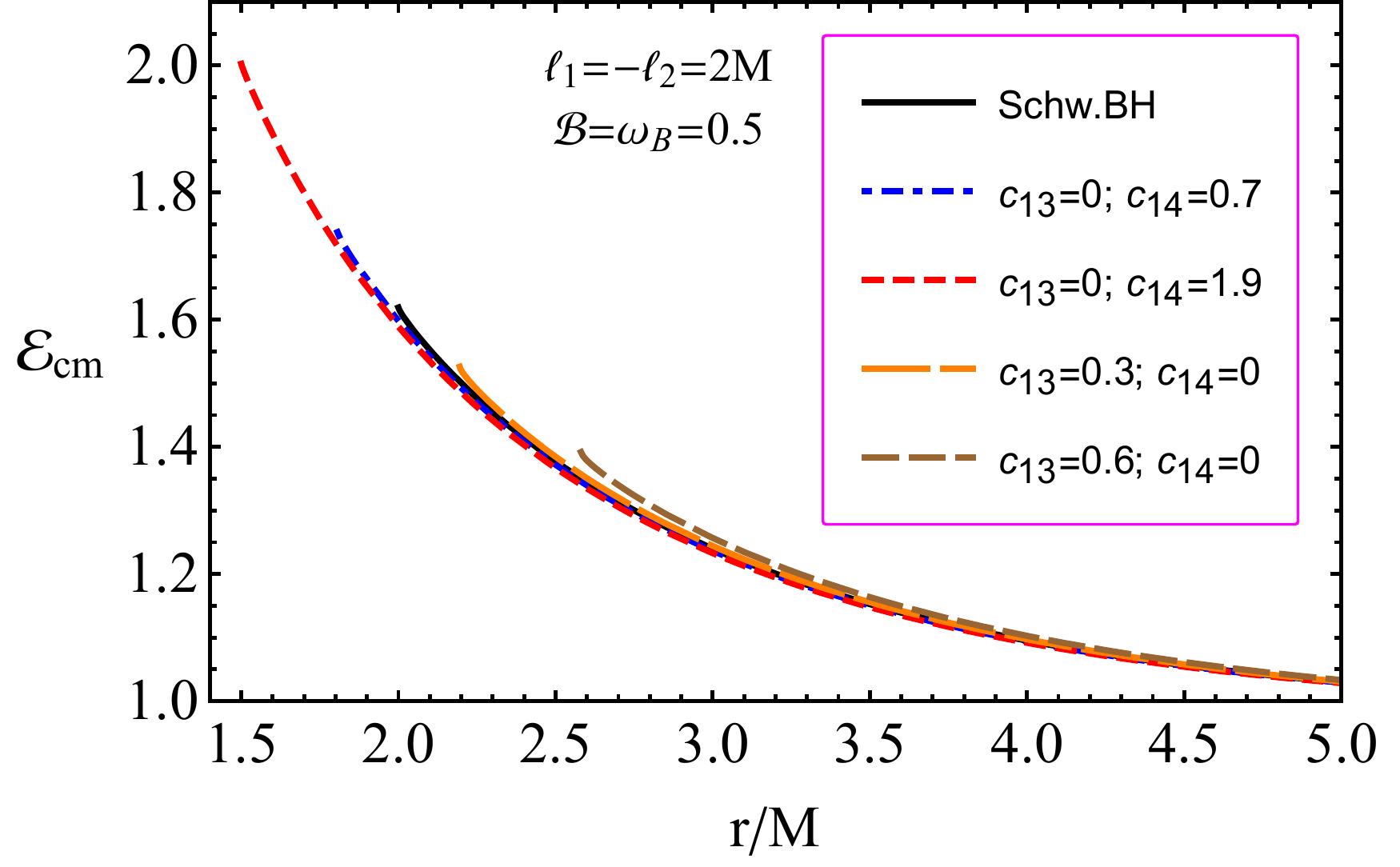}
   \includegraphics[width=0.98\linewidth]{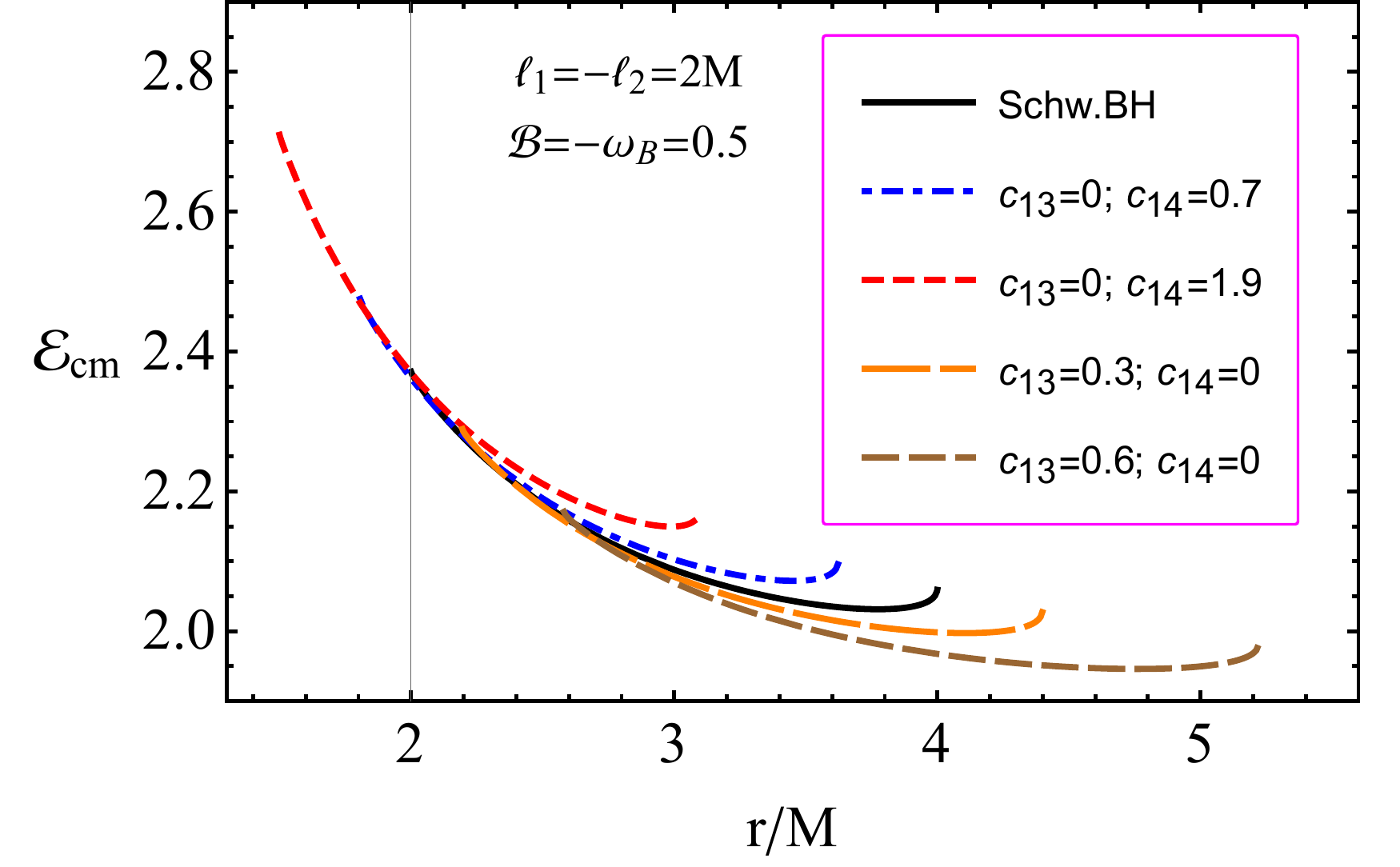}
       \caption{The radial dependence of the center-of-mass energy of the collisions of positively (on the top panel) and negatively (on the bottom panel) charged particles with magnetized particles for the different values of the aether gravity with comparisons of the Schwarzschild case. The coupling parameter of the magnetized particle is ${\cal B}=0.5$ while the cyclotron frequency for the charged particle is $\omega_B=0.5$. The angular momenta of the particles are $l_1=2M$ and $l_2=-2M$.} 
    \label{centermm2}
\end{figure}

The dependence of center-of-mass energy of the colliding electrically (positive and negative) charged particles and magnetized particles with the specific angular momentum $l_1=2M; l_2=-2M$ around the Einstein-\AE ther black hole along the radial coordinate is depicted in Fig.\ref{centermm2} for the different values of the parameters $c_{13}$ and $c_{14}$. One may see from the figures similar effects of the \AE ther parameters on the center-of-mass energy with the case of the collision of magnetized particles due to similar behaviour of magnetic interactions between the external magnetic field and magnetized (charged) particles. In the bottom panel of Fig.~\ref{centermm2} one can also see the cut off the center of mass energy similar as in Fig.~\ref{centermm}. In this scenario the particle's collision doesn't occur due to attractive effect of the external magnetic field on magnetized and charged particles. 

\subsection{Magnetized and neutral particles}

Finally, here we will carry on the studies of the collision of magnetized particles with neutral particles. One may immediately write the standard four-velocities for neutral particles in the spherical symmetric spacetime 

\begin{eqnarray}\label{4vel3}
\nonumber
\dot{t}&=&\frac{{\cal E}}{f(r)}\ ,
\\
\nonumber
\dot{r}^2&=&{\cal E}^2-f(r)\Bigg(1+\frac{l^2}{r^2}\Bigg)\ ,
\\
\dot{\phi}&=&\frac{l}{r^2}\ , 
\end{eqnarray}
and the expression for the center-of-mass energy of the collision

%%%%%%%%
\begin{eqnarray}\label{cmenergy3}
\nonumber
{\cal E}_{\rm cm}^2=1 &+& \frac{{\cal E}_1{\cal E}_2}{f(r)}-\frac{l_1l_2}{r^2}
-\frac{1}{f(r)}\sqrt{{\cal E}_2^2-f(r)\left(1+\frac{l_2^2}{r^2}\right)}
\\
& \times & \sqrt{{\cal E}_1^2-f(r)\left[\left(1-{\cal B} \sqrt{f(r)} \right)^2+\frac{l_1^2}{r^2} \right]}\ .
\end{eqnarray}

%%%%%%%
\begin{figure}
    \centering
   \includegraphics[width=0.98\linewidth]{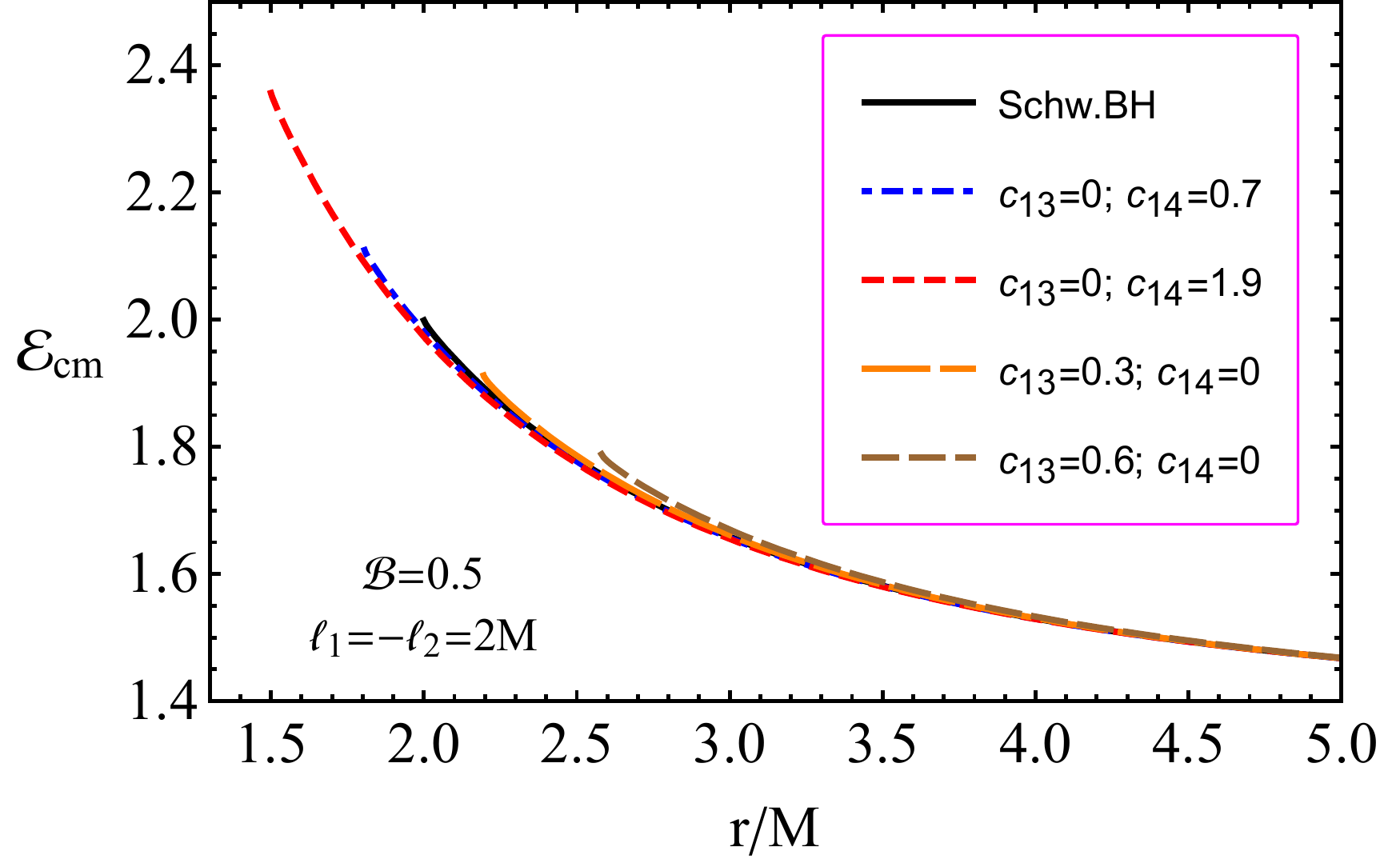}
       \caption{The radial dependence of the center-of-mass energy of  neutral and magnetized particles having the coupling parameter ${\cal B}=0.5$ with the angular momentum $l_1=2M$ and $l_2=-2M$. }
    \label{centermm3}
\end{figure}

Figure~\ref{centermm3} illustrates the radial dependence of center-of-mass energy of colliding magnetized and neutral particles for the different values of the \AE ther parameters. In these plots we have taken the values of the magnetic coupling parameter as ${\cal B}=0.5$ and considered head-on collision with the specific angular momentum $l_1=-l_2=2M$. One may conclude from Figs.~\ref{centermm}, \ref{centermm2} and \ref{centermm3} that the increase of the parameter $c_{14}$ ($c_{13}$) leads to increase (decrease) the center-of-mass energy in all collisions which we considered above.

\section{Static \AE ther black hole versus  Kerr black hole\label{appl}}

By considering the particle motion around black holes, one can test the properties of different theories of gravity. However, in most cases, the effects of different theories on measurable parameters of the central black hole may overlap each other and it becomes difficult to say the observed effect belongs to which particular theory or model. Indeed, there are many parameters in different gravity theories providing the exact same observational orbital parameters of the particle. Most astrophysical black holes are described as Kerr black hole. Here we test if the Einstein-\AE ther black hole immersed in an external asymptotically uniform magnetic field can mimic the rotation of the Kerr black hole by considering the magnetized particle motion and innermost stable circular orbits (ISCO). Particularly we consider magnetized particles motion around: 
\begin{enumerate}
\item Kerr black hole,
\item Einstein-\AE ther black hole,
\item Einstein-\AE ther black hole immersed in the magnetic field,
\item Schwarzschild MOG black hole.
\end{enumerate}  
ISCO radius for test particles around rotating Kerr black holes can be written in the following form 
\begin{eqnarray}\label{iscokerr}
r_{\rm isco}= 3 + Z_2 \pm \sqrt{(3- Z_1)(3+ Z_1 +2 Z_2 )} \ ,
\end{eqnarray}
with
\begin{eqnarray} \nonumber
Z_1 &  = &
1+\left( \sqrt[3]{1+a_*}+ \sqrt[3]{1-a_*} \right)
\sqrt[3]{1-a_*^2} \, ,
\\ \nonumber
Z_2 & = & \sqrt{3 a_*^2 + Z_1^2} , \qquad a_*=a/M \ ,
\end{eqnarray}
where $\pm$ corresponds to prograde and retrograde orbits. One may find the ISCO radius of the magnetized particle using the standard conditions
\begin{eqnarray}\label{iscocond}
\dot{r}=0\ , \qquad \frac{\partial V_{\rm eff}}{\partial r} =0\ ,\qquad {\rm and} \qquad \frac{\partial^2 V_{\rm eff}}{\partial r^2} = 0 \ , 
\end{eqnarray}
 where prime $'$ implies partial derivative with respect to  radial coordinate, where $V_{\rm eff}$ is the effective potential for radial motion of the magnetized particle in ZAMO frame which is defined by Eqs.(\ref{4vel1}) and (\ref{rdot}) in the following form
\begin{equation}
V_{\rm eff}(r)=f(r)\left[\left(1-{\cal B} \sqrt{f(r)} \right)^2+\frac{l^2}{r^2}\right]\ .
\end{equation}

%%%%%%%
\begin{figure}
    \centering
   \includegraphics[width=0.98\linewidth]{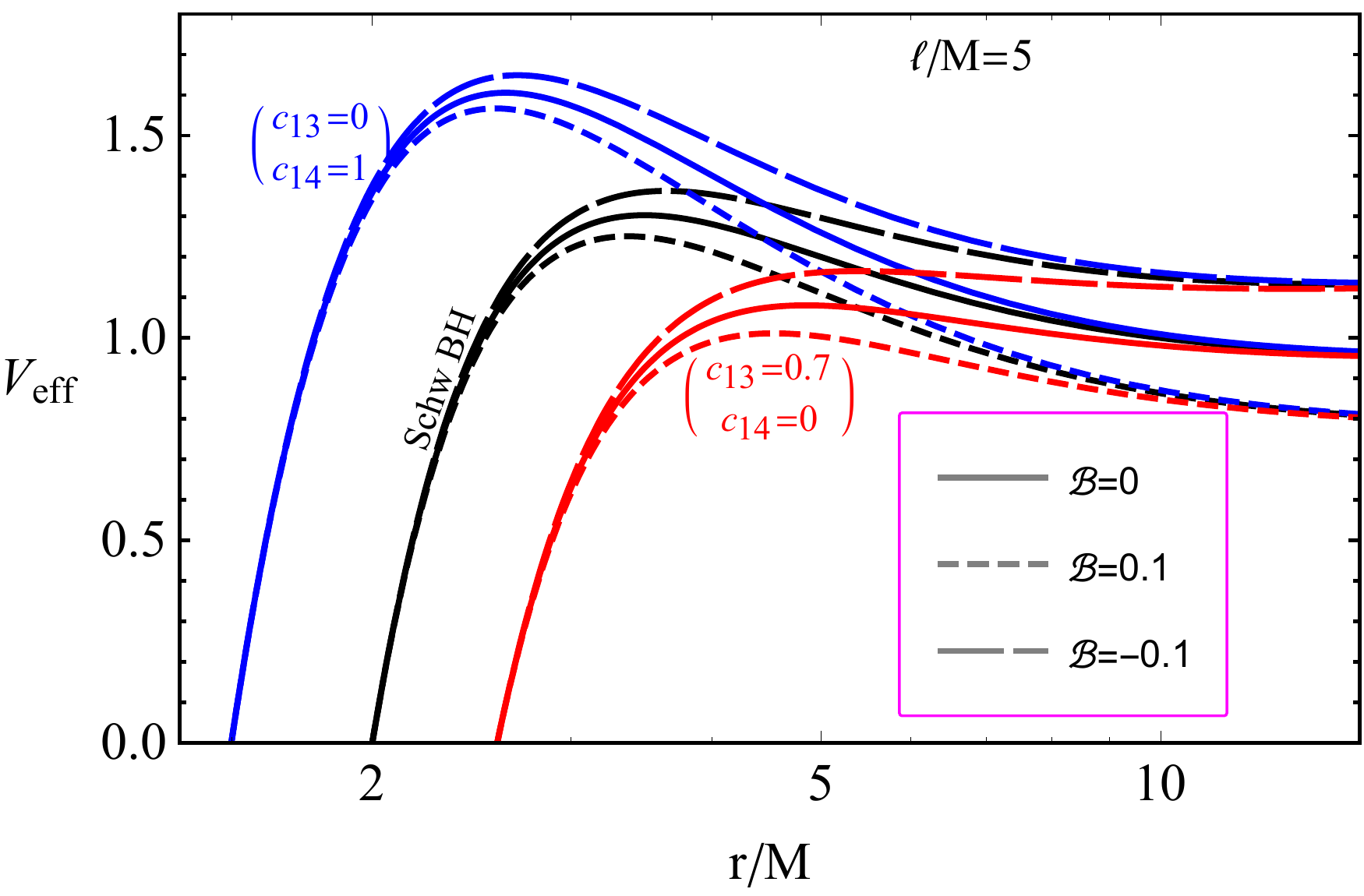}
       \caption{The radial dependence of effective potential for magnetized particles around static \AE ther black hole for different parameters of the black hole. Here, black, blue and red colors correspond to the cases ($c_{13}=c_{14}=0$), $c_{13}=0; c_{14}=1$ and $c_{13}=0.7; c_{14}=0$, respectively. Moreover, solid, large dashed and dashed lines correspond to zero, positive and negative values of the magnetic coupling parameter, respectively.}
    \label{effectivepot}
\end{figure}
Radial profiles of the effective potential of magnetized and neutral particles are shown in Fig.~\ref{effectivepot} for different values of the \AE ther parameters and compared with the Schwarzschild black hole case.
 One can see from Fig.\ref{effectivepot} that due to $c_{14}$ and negative magnetic coupling parameter lead to increase in the maximum of the effective potential while the parameter $c_{13}$ and positive magnetic coupling parameters causes to decrease the effective potential. Moreover, at far the effects of the \AE ther gravity disappear and magnetic interactions dominates. 

\begin{figure}
    \centering
   \includegraphics[width=0.9\linewidth]{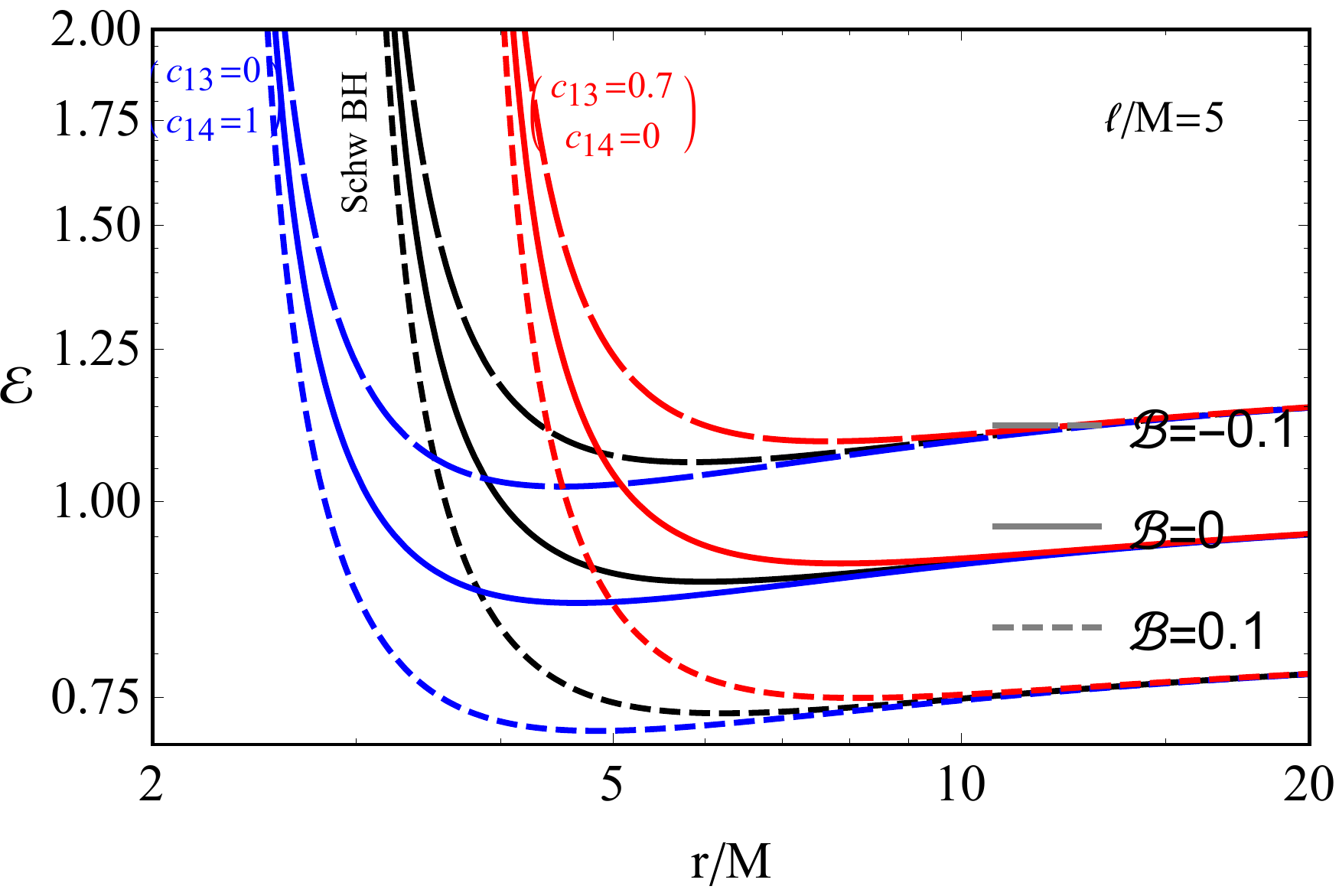}
   \includegraphics[width=0.98\linewidth]{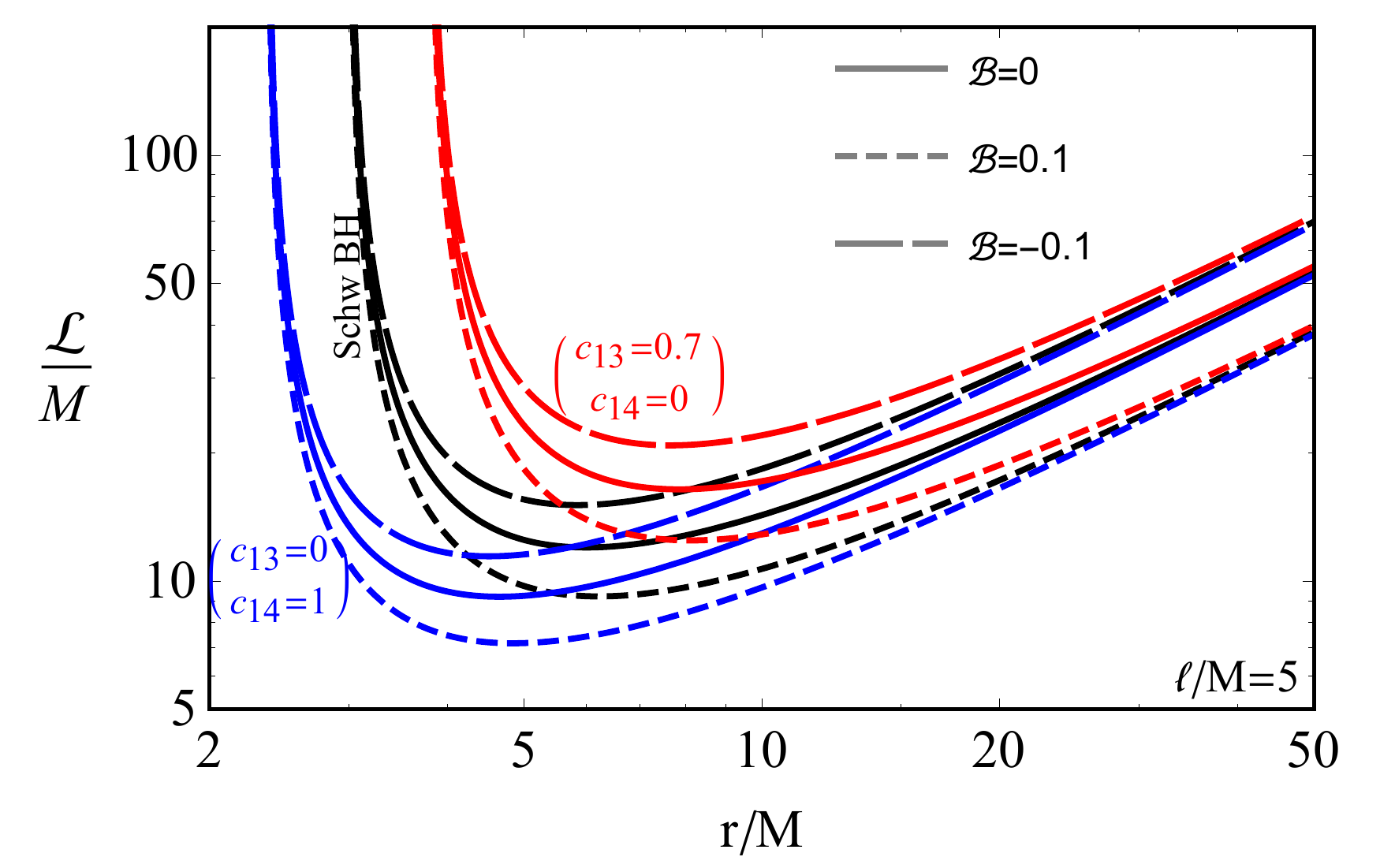}
       \caption{The radial dependence of specific angular momentum and energy of magnetized particles around static \ae ther black hole immersed in external asymptotically uniform magnetic field for the different parameters of the parameter of \ae ther gravity providing comparisons with Schwarzschild black hole.}
    \label{LLandEE}
\end{figure}

Now, we find the specific angular momentum and energy of the magnetized particle with the parameter ${\cal B}$, for circular motion in ZAMO frame using the second condition given in Eq.(\ref{iscocond}) in the following form
\begin{eqnarray}\label{llcritic}
    l^2&=&\frac{r^3 f'(r)}{2 f(r)-r f'(r)}\left[1-3 \mathcal{B} \sqrt{f(r)}+2 \mathcal{B}^2 f(r)\right] ,\\\nonumber
    {\cal E}&=&\frac{f(r)^{3/2} \left(\mathcal{B} \sqrt{f(r)}-1\right)}{2 f(r)-r f'(r)}\\& \times & \left[r \mathcal{B} f'(r)+2 \mathcal{B} f(r)-2 \sqrt{f(r)}\right].
\end{eqnarray}
Here we will provide graphical analysis of effects of the \ae ther gravity parameters on specific angular momentum and energy of magnetized particles with the magnetic coupling parameter $|{\cal B}|=0.1$ in Fig.\ref{LLandEE}

 One can see from Fig.\ref{LLandEE} that the minimal values of the specific angular momentum and energy increases with the increase of $c_{13}$ and negative values of the magnetic coupling parameter. However, by increasing $c_{14}$ and considering positive magnetic field decreases the minimum of the energy and angular momentum. Moreover, the minimum radius of circular orbits increases (decreases) with the increase of $c_{13}$ ($c_{14}$) due to increase of gravitational potential of central object.
 
\subsection{Trajectories of magnetized particles}

In this subsection we will study bound orbits of a magnetized particle around black hole immersed in an external magnetic field in the presence of aether. We consider the separate cases when $c_{13}=0$ and $c_{14}=0$, respectively. Here we have chosen the magnetic coupling parameter as ${\cal B}=0.5$ and specific angular momentum of the particle $l/M=2\sqrt{5}$.

\begin{figure*}[ht!]
  \centering
   \includegraphics[width=0.98\linewidth]{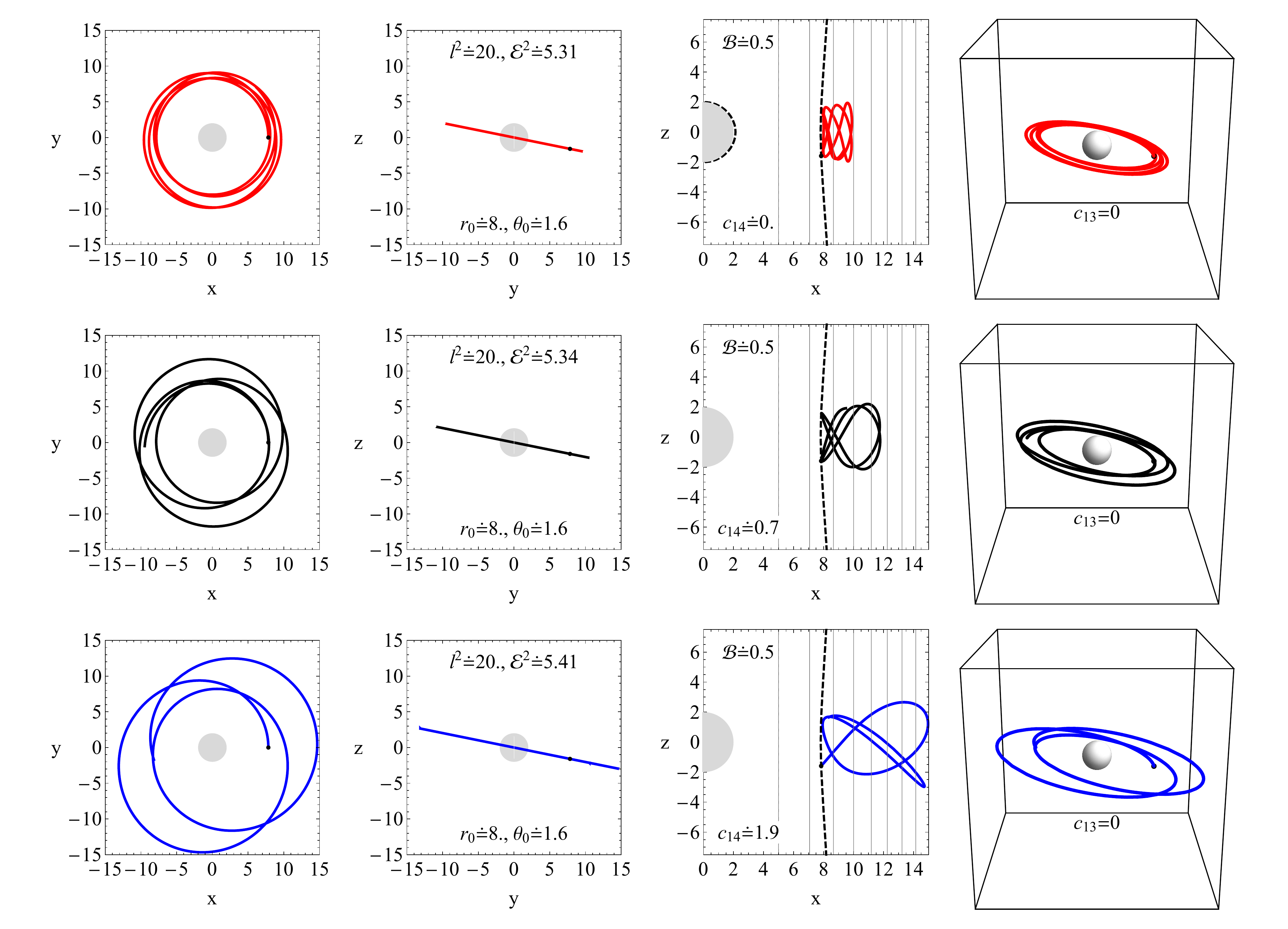}
    \caption{Trajectory profiles of the magnetized particle with the magnetic coupling parameter ${\cal B}=0.5$, for different values of $c_{14}$ and the fixed specific angular momentum $l^2/M^2=20$ when $c_{13}=0$.     \label{tr1}}
\end{figure*}

\begin{figure*}[ht!]
  \centering
      \includegraphics[width=0.98\linewidth]{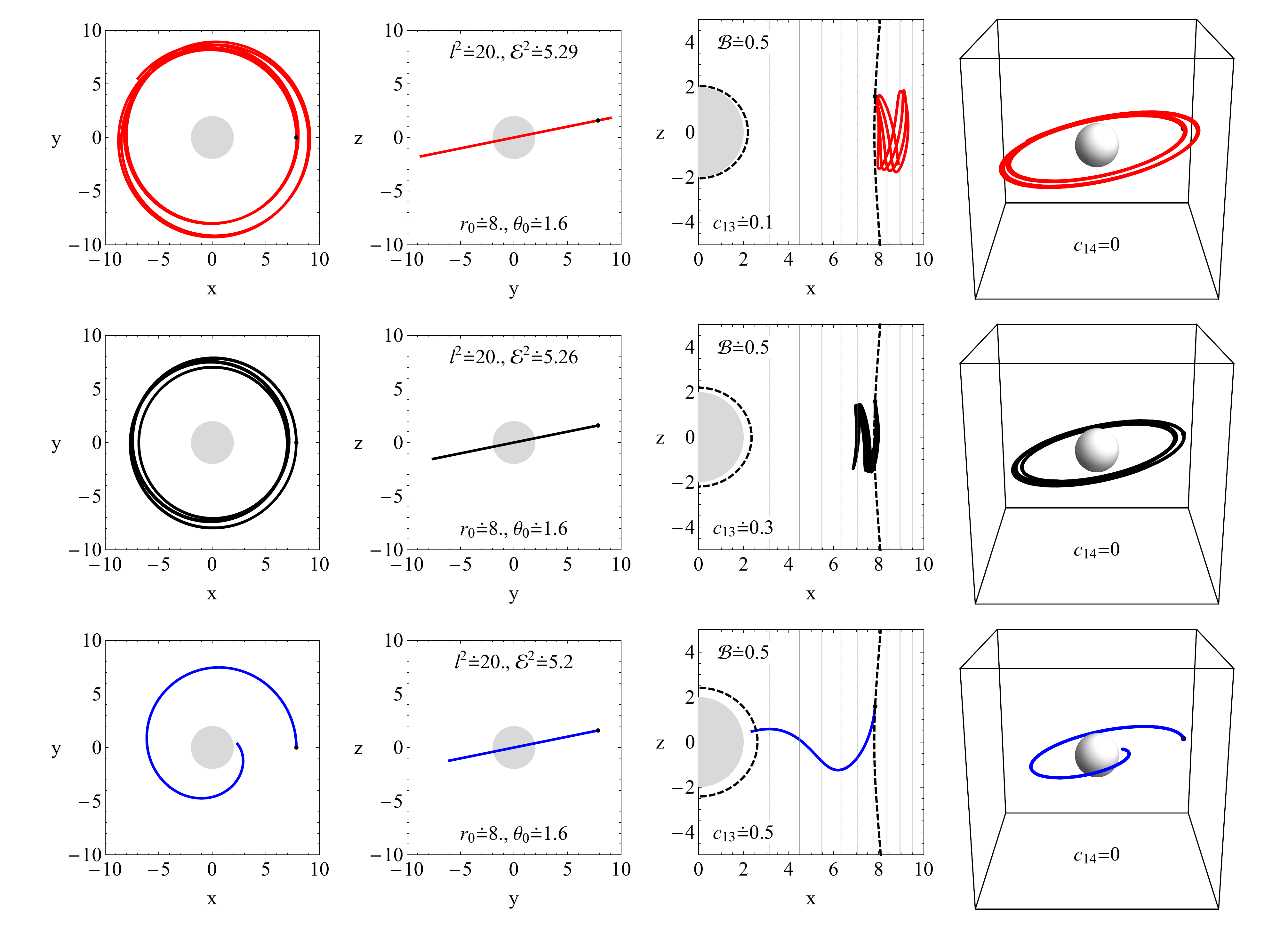}
    \caption{The same figure with Fig.~\ref{tr1}, but for the case when $c_{14}=0$.     \label{tr2}}
\end{figure*}

Figures~\ref{tr1} and \ref{tr2} depict trajectories of the magnetized particle with the coupling parameter ${\cal B}=0.5$ around the black hole in Einstein-\AE ther gravity for the cases $c_{13}=0$ and $c_{14}=0$, respectively.  For plotting the trajectories, we select initial position of the particles as $r_0=8M$ and $\theta_0=\pi/2$ with the fixed time duration. One can see from the trajectories in Figs.~\ref{tr1} and \ref{tr2} that the bound orbits are larger in the absence of the parameter $c_{13}$ than the case of vanishing parameter $c_{14}$. Moreover, the increase of the parameter $c_{14}$ ($c_{13}$) causes the increase (decrease) in the radius of bounded orbits.  The energy of the particle along the orbits increases (decreases) with the increase of the parameter $c_{14}$ ($c_{13}$). With the increase of  $c_{13}$ the bounded orbits becomes unstable and the particles fall down to the central object. One may conclude from the comparisons of the orbits that by increasing the parameter $c_{13}$ ($c_{14}$) leads to increase (decrease) of the gravitational potential of the central object.

One may easily find equation for the radius of ISCO putting the specific angular momentum given in Eq. (\ref{llcritic}) to the last condition given in Eq.(\ref{iscocond}) and we have 
\begin{eqnarray}\label{ISCOb}
\nonumber
&&f'(r) \Big[r^2 \mathcal{B} \left(4 \mathcal{B} \sqrt{f(r)}-3\right) f'(r)^2 \\\nonumber &&+2 r \sqrt{f(r)} \left(4 \mathcal{B}^2 f(r)-9 \mathcal{B} \sqrt{f(r)}+4\right) f'(r)\\ &&-12 f(r)^{3/2} \left(2 \mathcal{B}^2 f(r)-3 \mathcal{B} \sqrt{f(r)}+1\right)\Big]\\\nonumber &&-4 r f(r)^{3/2} \left(2 \mathcal{B}^2 f(r)-3 \mathcal{B} \sqrt{f(r)}+1\right) f''(r)\geq 0 \ .
\end{eqnarray}

 One can see that obtaining the exact analytical solution of Eq.(\ref{ISCOb}) is impossible. However, we provide detailed numerical analysis for the cases when the ISCO radius of the magnetized particle with magnetic coupling parameters ${\cal B}=0.5$ and  ${\cal B}=-0.5$ around Einstein-\AE ther and Kerr black holes take the same values, i.e. those values of the \AE ther and spin parameters which provide exactly the same values for ISCO radius of magnetized particles, using Eqs.(\ref{iscokerr}) and (\ref{ISCOb}) for the following cases.

\subsection{$c_{14}=0$}
 
\begin{figure*}
    \centering
   \includegraphics[width=0.7\linewidth]{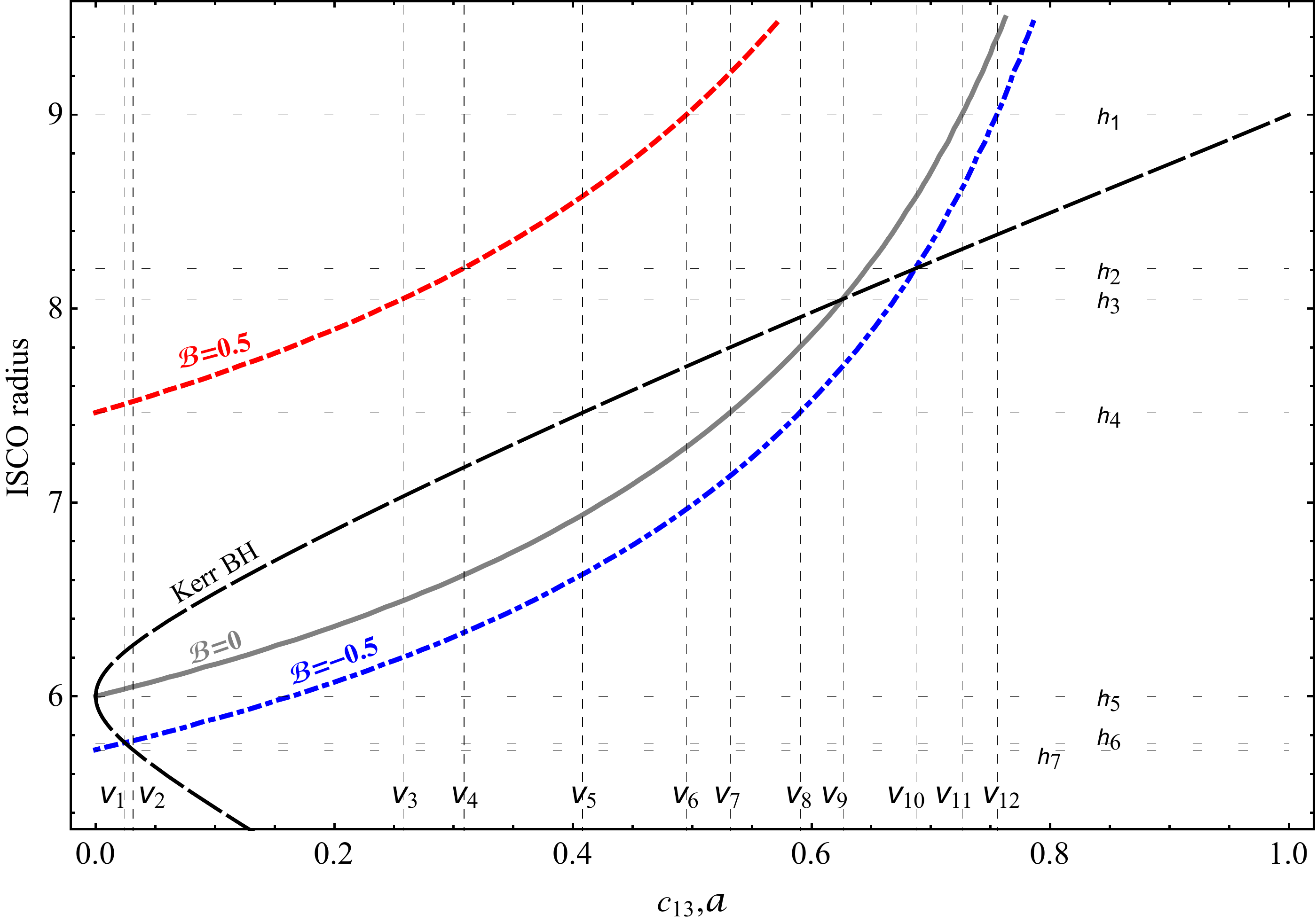}
       \caption{The dependence of ISCO radius on spin and Einstein-\AE ther parameter $c_{13}$ when $c_{14}=0$. Black large dashed and grey solid lines correspond to the ISCO of magnetized particles around Kerr and the Einstein-\AE ther black holes in the absence of the magnetic field, respectively. Blue dot-dashed and red dashed lines correspond to ISCO of the magnetized particle around the Einstein-\AE ther black hole immersed in the magnetic field for the values of the magnetic coupling parameter ${\cal B}_1=-0.5$ and ${\cal B}_2=0.5$, respectively. The black large dashed line corresponds to ISCO radius as a function of the spin parameter; red, blue, and grey lines correspond to ISCO radius as a function of the parameter $c_{13}$ for different values of ${\cal B}$. Vertical (${\rm v}_i,\ i=1-12$) and horizontal (${\rm h}_j,\ j=1-7$) lines imply the important values for the spin of Kerr and Einstein-\AE ther black hole parameters and the values of the ISCO radius where the lines intersect (see the text for discussion). }
    \label{c13vsa}
\end{figure*}

ISCO profiles of the magnetized particle around Kerr and Einstein-\AE ther black holes in the presence and absence of the external magnetic field for the case of when the parameter $c_{14}=0$ are shown in Fig.~\ref{c13vsa}. One can see that the parameter $c_{13}$ can mimic the innermost counter rotating orbits of the particle around Kerr black hole up to $c_{13}=0.7556$ (${\rm v}_{12}$ vertical line) for ${\cal B}=-0.5$,  $c_{13}=0.7262$ (${\rm v}_{11}$ vertical line) for ${\cal B}=0$ and  $c_{13}=0.4954$ (${\rm v}_{6}$ vertical line) for ${\cal B}=0.5$ in the range of ISCO radius $5.7243 \leq r_{\rm ISCO}/M\leq 9$ (${\rm h}_{1}$ and ${\rm h}_{7}$ horizontal lines). 

ISCO radius is the same when the values of the spin and the Einstein-\AE ther parameter $c_{13}=a/M=0.6266$ (${\rm v}_{9}$ vertical line) in the absence of the external magnetic field, $c_{13}=a/M=0.6875$ (${\rm v}_{10}$ vertical line) for ${\cal B}=-0.5$ and $c_{13}=a/M=0.0244$ (${\rm v}_{1}$ vertical line) at $r_{\rm ISCO}/M=8.0501$ ( ${\rm h}_{3}$ horizontal line), $r_{\rm ISCO}/M=8.2077$ (${\rm h}_{2}$ horizontal line) and $r_{\rm ISCO}/M=5.7605$ ( ${\rm h}_{6}$ horizontal line), respectively.

ISCO radius of the magnetized particle with the coupling parameter ${\cal B}=0.5$ around Einstein-\AE ther black hole in the presence of the external magnetic field can be the same when it is around Kerr black hole in the absence of the magnetic field provided the spin parameter $0.4081\leq a/M \leq 1$ (${\rm v}_{5}$ vertical line) with the radius $7.463 \leq r_{\rm ISCO}/M \leq 9$ (${\rm h}_{1}$ and ${\rm h}_{4}$ horizontal lines).

Moreover, one can observe the effect of the external magnetic field looking at the red, blue and grey lines from Fig.~\ref{c13vsa}. One can see that it is not possible to distinguish the existence of the external magnetic field at $c_{13}\geq 0.5314$ (${\rm v}_{5}$ vertical line) because of the ISCO is the same for the particle with ${\cal B}=0.5$. One may distinguish the orientation of the external magnetic field with respect to the direction of the magnetized particle's dipole moment at $c_{13}<0.5906$ (${\rm v}_{8}$ vertical line) and $r_{\rm ISCO}/M> 7.463$.

\subsection{$c_{13}=0$}
 
\begin{figure*}
    \centering
   \includegraphics[width=0.7\linewidth]{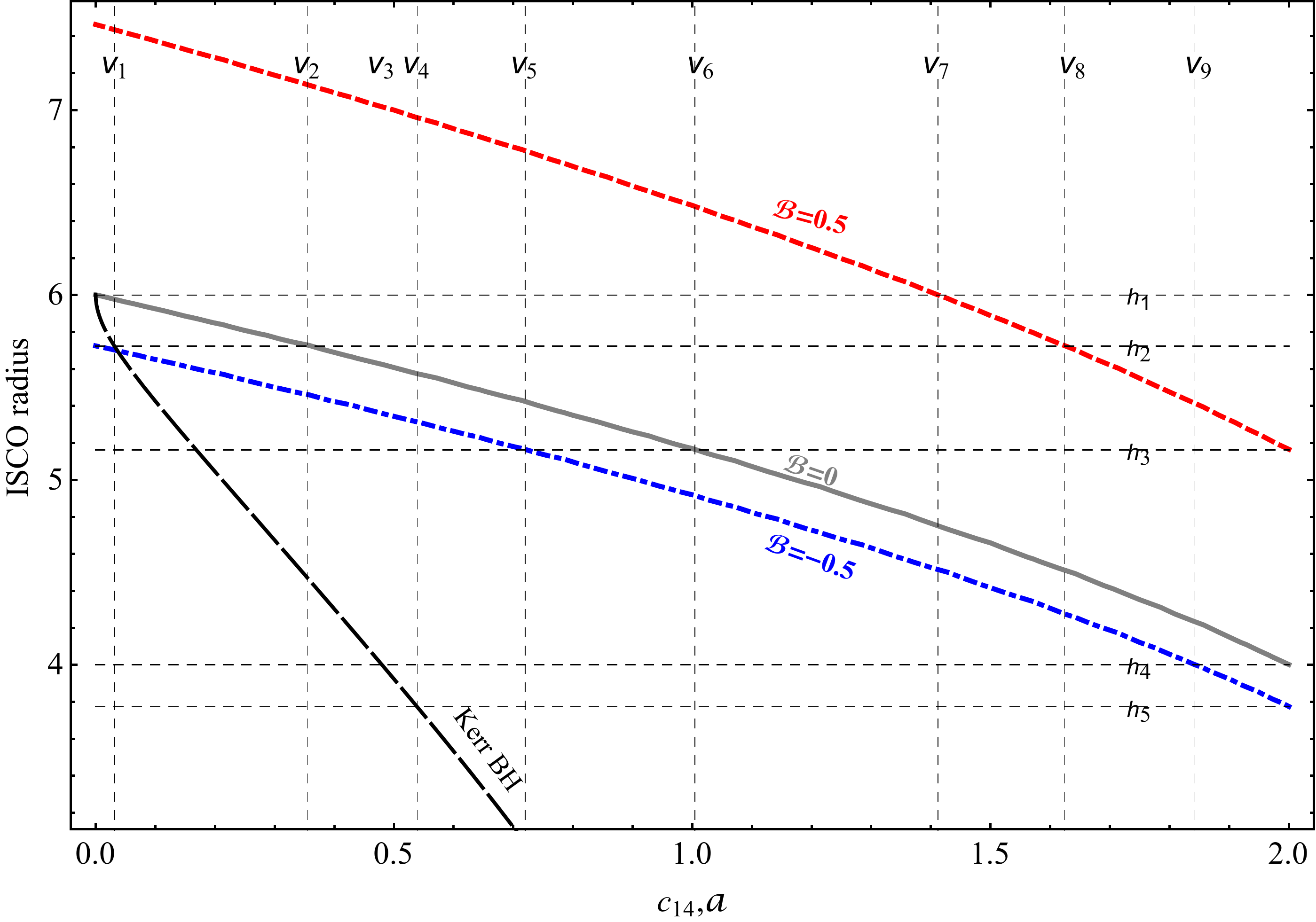}
       \caption{The same figure as Fig.~\ref{c13vsa}, but for the case $c_{13}=0$.}
    \label{c14vsa}
\end{figure*}

In this subsection, we make similar discussions for the case of $c_{13}=0$.  It is not difficult to see that Fig.~\ref{c14vsa} is similar to Fig.~\ref{c13vsa} just for the Einstein-\AE ther parameter $c_{13}=0$. One can see from the figure that in this case the effect of the another Einstein-\AE ther gravity parameter $c_{14}$ can mimic the spin of the Kerr black hole giving the same innermost co-rotating orbits.

One can see that in the cases, when the external magnetic field is present at the values of the spin parameter $a/M>0.5388$ (${\rm v}_1$ vertical line) and $a/M<0.0312$ (${\rm v}_4$ vertical line), can not mimic the Einstein-\AE ther gravity parameter for the magnetized particle with the coupling parameter ${\cal B}=-0.5$. The parameter $c_{14}$ mimic spin of Kerr black hole in the range $6>r_{\rm isco}/M>5.1623$ (${\rm h}_{1}$ and ${\rm h}_{3}$ horizontal lines) at $2\geq c_{14}\geq 1.4119$  (${\rm v}_{7}$ vertical line) for the magnetized particle with the parameter ${\cal B}=0.5$. Further in the absence of the external magnetic field, the spin parameter can not mimic the parameter $c_{14}$ when $a/M\geq 0.4799$ (${\rm v}_{3}$ vertical line). 

Now again looking for the effect of magnetic field, we will see at red, blue and grey lines. One can notice that the magnetic interaction can not mimic the effect of the parameter at the the range $c_{14}\leq 1.4119$  (${\rm v}_{7}$ vertical line) for ${\cal B}=0.5$ and $c_{14} \geq 1.8426$ for  ${\cal B}=-0.5$  (${\rm v}_{9}$ vertical line).

The orientation of the external magnetic field can be distinguishable when the Einstein-\AE ther gravity parameter $c_{14}<1.6241$ (${\rm v}_{8}$ vertical line)  for ${\cal B}=0.5$ and $c_{5}>0.72$ for ${\cal B}=-0.5$ in the range of ISCO radius $5.7243 \leq r_{\rm ISCO}/M \leq 5.1623$ (${\rm h}_{2}$ and ${\rm h}_{3}$ horizontal lines).

\subsection{$c_{13=0}$ solution vs $c_{14}=0$ one}

In this subsection we analyse ISCO radius of magnatized particles with the magnetic coupling parameter ${\cal B}=0.5$ and ${\cal B}=-0.5$ around Einstein-\AE ther black hole in the presence and absence of the external parameters that in with cases the solution $c_{14}=0$ can mimic the solution $c_{13}=0$.

\begin{figure*}
    \centering
   \includegraphics[width=0.7\linewidth]{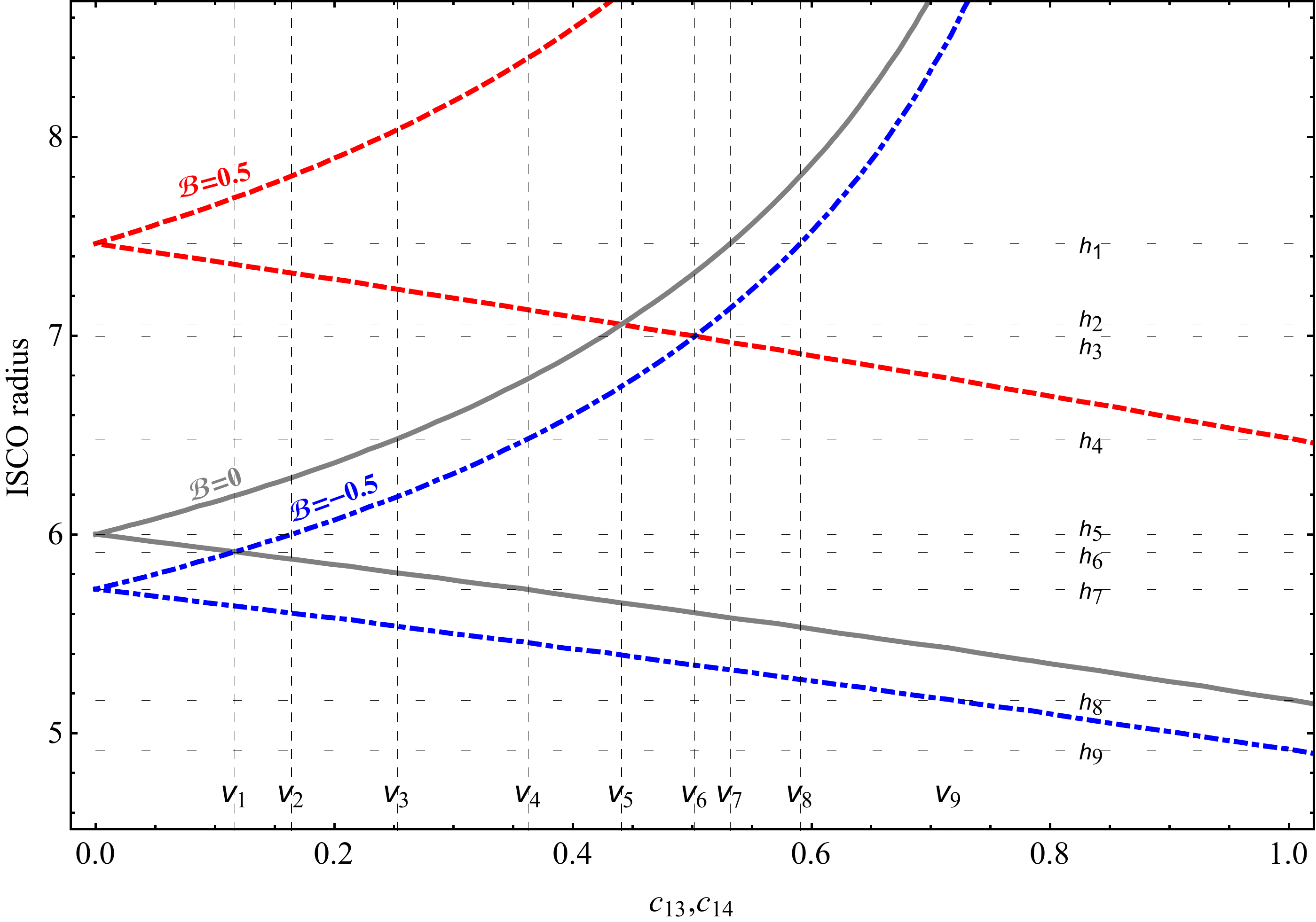}
       \caption{The same comparisons with Figs.~\ref{c13vsa} and \ref{c14vsa} but for the two Einstein-\ae ther parameters $c_{13}$ and $c_{14}$.}
    \label{c13vs14}
\end{figure*}

Figure~\ref{c13vs14} illustrates ISCO profiles of the magnatized particle around Einstein-\AE ther black holes with solutions $c_{13}=0$ and $c_{14}=0$ in the presence $|{\cal B}|=0.5$ and absence of the external magnetic field. One can see from the figure that ISCO radius of the particle is same in the cases: {\bf a}) the particle with coupling parameter ${\cal B}=0.5$ and  ${\cal B}=-0.5$ at $c_{13}=c_{14}=0.502$ (${\rm v}_{6}$ vertical line)  with the radius $r_{\rm isco}/M=6.9954$ (${\rm h}_{3}$  horizontal line); {\bf b}) ${\cal B}=0.5$ and  ${\cal B}=0$ at $c_{13}=c_{14}=0.4407$ (${\rm v}_{5}$ vertical line)  with the radius $r_{\rm ISCO}/M=7.0549$ (${\rm h}_{2}$  horizontal line); {\bf c}) ${\cal B}=-0.5$ and  ${\cal B}=0$ at $c_{13}=c_{14}=0.1166$ (${\rm v}_{1}$ vertical line)  with the radius $r_{\rm ISCO}/M=5.9116$ (${\rm h}_{6}$  horizontal line).

ISCO radius of the magnatized particle with the magnetic coupling parameter ${\cal B}=0.5$ and  the parameter $c_{14} \in (0,1)$ can be measured the same as the ISCO of the particle with ${\cal B}=-0.5$ and ${\cal B}=0$ with the parameter $c_{13} \in (0.3626, 0.5905)$ (${\rm v}_{4}$ and ${\rm v}_{8}$ vertical lines) and $c_{13} \in (0.2528, 0.5318)$ (${\rm v}_{3}$ and ${\rm v}_{7}$ vertical lines) in the range of the radius $7.463 \geq r_{\rm ISCO}/M \geq 6.4809$ (${\rm h}_{1}$ and ${\rm h}_{4}$ horizontal lines). Moreover, ISCO radius of the particle with ${\cal B}=0$, at the values of the parameter $c_{14} \in (0,0.3625)$  (${\rm v}_{4}$ vertical line) can be the same with the ISCO radius of a magnetized parameter with ${\cal B}=-0.5$,  at the values of the parameter $c_{13} \in (0,0.1641)$ (${\rm v}_{2}$ vertical line) for the range of ISCO radius of $6 \geq r_{\rm ISCO}/M \geq 5.7243$ (${\rm h}_{5}$ and ${\rm h}_{7}$ horizontal lines). 

\subsection{Schwarzschild MOG black hole versus \AE ther black hole}

In this subsection we compare the effects of MOG and \AE ther parameter on magnetized particles dynamics. The spacetime metric around static black hole in modified gravity can be described by the following lapse function 
\begin{equation}
    f(r)=1-\frac{2(1+\alpha)G_N M}{r}+\frac{\alpha (1+\alpha)G_N^2 M^2}{r^2}\ .
\end{equation}

The studies of dynamics of magnetized particles around Schwarzschild MOG black holes immersed in an external asymptotically uniform magnetic fields has been performed in our previous paper \cite{Haydarov2020EPJC} and was shown how the static MOG black hole can reflect the effects of spin of Kerr black holes proving exactly the same values of ISCO radius of magnetized particles. 
Now, here, we will be interested to answer whether a Schwarzschild BH in MOG covers the effects of the black hole in Einstein-\AE ther gravity on the measurements of the ISCO radius of magnetized particles in the presence of the external magnetic fields, in the other words can \AE ther black hole be described by MOG with suitable values of the MOG parameter. Here we will make similar analysis on the effects of both MOG and \AE ther parameters on ISCO radius of magnetized particles with the magnetic coupling parameters ${\cal B}=0.1$.  

\begin{figure}
    \centering
   \includegraphics[width=0.99\linewidth]{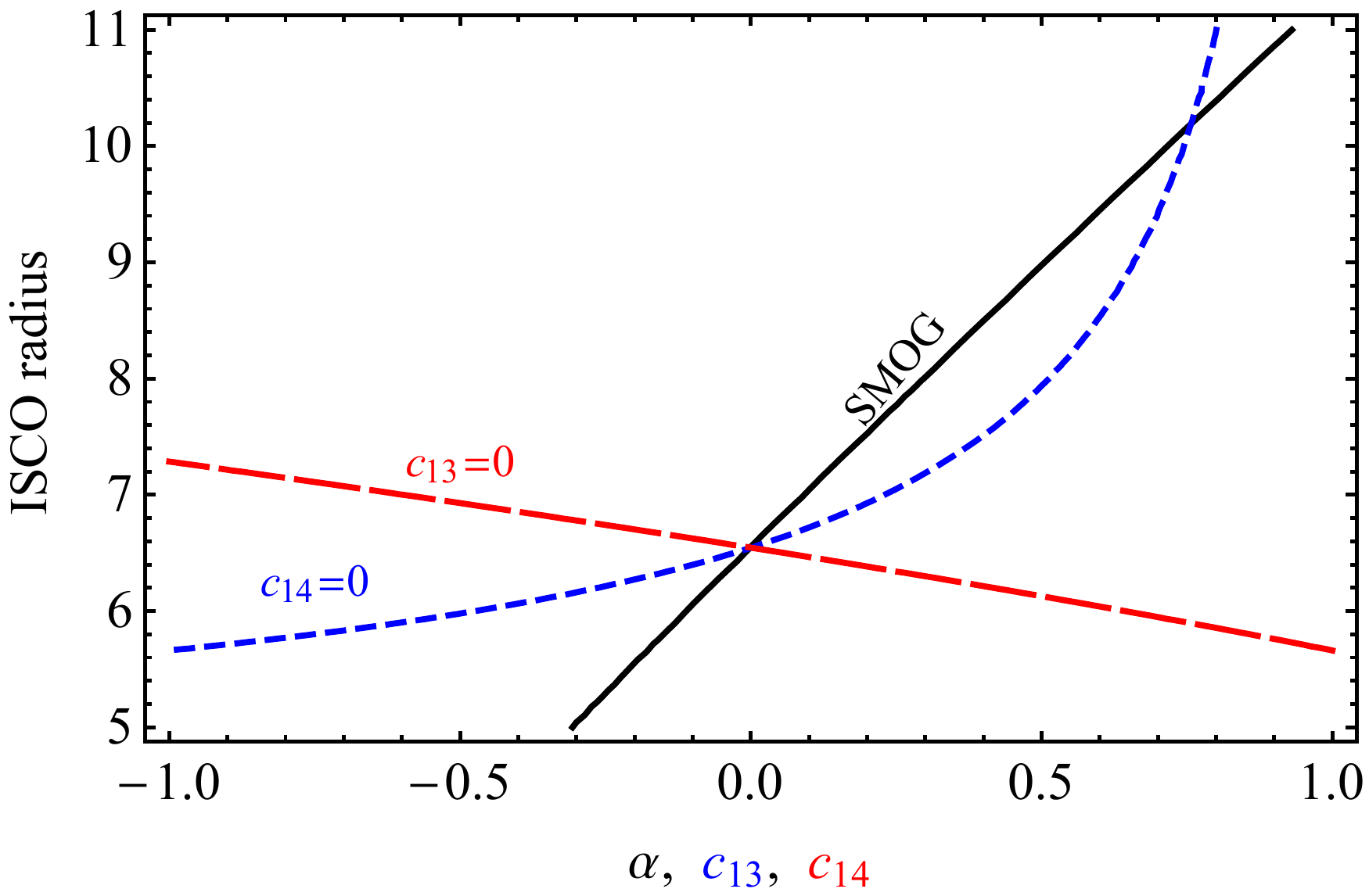}
       \caption{Dependence of ISCO radius of the magnetized particle with the magnetic coupling parameter ${\cal B}=0.1$ around a static black hole from the two Einstein-\ae ther gravity ($c_{13}$,$c_{14}$) and MOG parameters ($\alpha$) .}
    \label{mogvsae}
\end{figure}

Figure \ref{mogvsae} illustrates the effects of Einstein-\AE ther and modified gravity parameters on ISCO radius of magnetized particles. Here we analyse effects of the \AE ther gravity in two cases: (i) $c_{13}=0$ and (ii) $c_{14}=0$ and we have tested all the parameters in the range of $(-1,\ 1)$. One can see that in case when $c_{13}=0$, effects of the parameters $\alpha$ and $c_{14}$ can cover each other giving exactly the same ISCO radius in the range $r_{\rm ISCO}/M\in (5.6625;  \ 7.28609)$ for the magnetar with the magnetic coupling parameter ${\cal B}=0$, in the range of their values $\alpha \in (-0.17654; \ 0.150636)$ and $c_{14} \in (-1,\ 1)$, while in the case when $c_{14}=0$ the mimic values lies $\alpha \in (-0.17654, \ 0.755868)$ and $c_{13}\in (-1, \ 0.755868)$ for the ISCO radius $r_{\rm ISCO}/M \in (5.6625, \ 10.1804)$. In the other words, when the measured ISCO radius of a magnetized particle lies out of the range one cannot distinguish it whether it is due to effects of SMOG or \AE ther.

\section{Summary and Discussions\label{Summary}}

This work is devoted to study the effects of Einstein-\AE ther gravity on magnetized particle motion around a static and spherically symmetric uncharged black hole immersed in an external asymptotically uniform magnetic field. The following major results have been obtained:

\begin{itemize}

\item The electromagnetic field solution has been obtained using Wald's method and shown that the existence of the parameter $c_{13}$ ($c_{14}$) causes the decrease (increase) of the magnetic field near the black hole.

\item The studies of the circular motion of the magnetized particles show that inner circular (bounded) orbits come closer to (or goes far from) the central object and the range where circular orbits are allowed increases (decreases) in the presence of the parameter $c_{14}$ ($c_{13}$). It implies that the parameter $c_{13}$ plays a role as an additional gravity effect (in other words, due to the existence of parameter $c_{13}$ the effective mass of central black hole increases).

\item The analysis of the head-on collisions of magnetized particles around the \AE ther black hole showed that the parameter $c_{13}$ ($c_{14}$) causes the increase (decrease) of center-of-mass energy and show that collisions of magnetized particles in the opposite direction of their magnetic dipoles do not occur due to dominating effect of attractive magnetic interaction. The distance where the collision of the magnetized particles does not occur goes far (or comes closer) due to an increase of the parameter $c_{13}$ ($c_{14}$).   

\item We have shown through analysis of the effective potential for a radial motion of the magnetized particles in ZAMO frame, that the maximum value of the effective potential increase (decreases) with the increase of the parameter $c_{14}$ ($c_{13}$ and negative magnetic coupling parameter).  

\item We have provided trajectories of magnetized particles around the \AE ther black holes in the external magnetic fields and shown that the increase of the parameter $c_{13}$ ($c_{14}$) cause the increase (decrease) of the radius of the stable orbits making stronger (weaker) the gravitational potential of the central black hole.

\item We have also analyzed the effects of Einstein-\AE ther gravity and magnetic interaction can jointly mimic the effects of the rotation of the Kerr black hole by giving the same ISCO radius. It was also shown that the magnetized particles may have the same ISCO radius around rotating Kerr and Einstein-\AE ther black holes at the range of $r_{\rm isco}/M \in (5.7243, 9)$ at the parameter $c_{13}\leq 0.7556$ (0.7262 and 0.4954) for the particles with ${\cal B}=0.5$ (${\cal B}=0.5$ and ${\cal B}=0$)  when $c_{14}=0$ and the spin parameter can not mimic the parameter $c_{14}$ when $a/M>0.4799$.

\item ISCO radius is exactly the same for the values of the spin and the Einstein-\AE ther parameter $c_{13}=a/M=0.6266$  in the absence of the external magnetic field, $c_{13}=a/M=0.6875$ for ${\cal B}=-0.5$ and $c_{13}=a/M=0.0244$ at $r_{\rm isco}=8.0501$ , $r_{\rm isco}/M=8.2077$  and $r_{\rm isco}/M=5.7605$, respectively.

\item ISCO radius of the particle is  exactly the same in cases of the magnetized particle with coupling parameter ${\cal B}=0.5$ and  ${\cal B}=-0.5$ at $c_{13}=c_{14}=0.502$ with the radius $r_{\rm isco}=6.9954$. For ${\cal B}=0.5$ and  ${\cal B}=0$ at $c_{13}=c_{14}=0.4407$  with the radius $r_{\rm isco}=7.0549$. For ${\cal B}=-0.5$ and  ${\cal B}=0$ at $c_{13}=c_{14}=0.1166$  with the radius $r_{\rm isco}=5.9116$.

\item Finally, we have compared effects of MOG parameter and parameters of Einstein- \AE ther gravity on the ISCO radius of the magnetized particles with the parameter ${\cal B}=0.2$ and shown that the MOG and $c_{14}$ parameters mimics each other for the ISCO radius $r_{\rm ISCO}/M\in (5.6625;  \ 7.28609)$ for $c_{13}=0$ case, while when $c_{14}=0$ ISCO radius is $r_{\rm ISCO}/M=10.1806 $ for the values of the parameters $c_{13}=\alpha=0.755868$.

\end{itemize}

\subsection*{Acknowledgement}

The research work of AA is supported by postdoc fund PIFI of Chinese academy of sciences.  This research is supported by Grants No. VA-FA-F-2-008, No.MRB-AN-2019-29 of the Uzbekistan Ministry for Innovative Development. JR and AA thank Silesian University in Opava for the hospitality during their visit.

\bibliographystyle{apsrev4-1}
\bibliography{gravreferences}

\end{document}